\documentclass[12pt]{article}
\usepackage[utf8]{inputenc}
\usepackage[numbers,sort&compress]{natbib}
\usepackage{geometry}
\usepackage{mathtools,mathrsfs}
\usepackage{dsfont,relsize}
\usepackage{newtxtext,newtxmath}
\usepackage{amsmath}
\usepackage{amsbsy}
\usepackage{cancel}
\usepackage{graphicx}
\usepackage{float}
\usepackage{ytableau}
\usepackage{color}
\usepackage{graphicx}
\usepackage{mathtools,slashed}
\renewcommand{\emph}[1]{\textit{#1}}
\usepackage{simplewick}

\usepackage{authblk}

\usepackage{xargs}
\newcommandx{\overbar}[1]{\mkern 5.mu\overline{\mkern-5.mu#1\mkern-1.0mu}\mkern 1.5mu}
\DeclareMathAlphabet{\mathpzc}{OT1}{pzc}{m}{it}

\definecolor{PineGreen}{cmyk}{0.92, 0, 0.59, 0.25}

\newcommand{\HC}{\mathcal{H}_{\text{C}}}
\usepackage{ifpdf}
\ifx\pdfoutput\undefined
   \pdffalse
   \usepackage{cite}
 \else
   \pdfoutput=1
   \pdftrue
  \usepackage[pdftex]{hyperref}
\fi

\usepackage{upgreek}
\numberwithin{equation}{section}
\allowdisplaybreaks
\usepackage[all]{xy}
\usepackage{color}
\usepackage{graphicx}
\graphicspath{{images/}}

\usepackage{newtxtext,newtxmath}

\usepackage[utf8]{inputenc}
\usepackage{ragged2e}
\usepackage{appendix}
\usepackage{slashed}

\usepackage{enumitem}

\usepackage{soul}

\usepackage{mathtools}

\title{\LARGE\bf Dynamical Implementation of the Constraints in Conformal Gravity}

\author{L. Andrianopoli$^{[a,b]}$, R. D'Auria$^{[a]}$, G. Grosso$^{[a]}$, L. Ravera$^{[a,b,c,*]}$}

\date{\today}

\begin{document}

\maketitle

\vskip 0.2cm

\noindent
$^{[a]}$ DISAT, Politecnico di Torino, Corso Duca degli Abruzzi 24, 10129 Torino, Italy. \\
$^{[b]}$ INFN, Sezione di Torino, Via Pietro Giuria 1, 10125 Torino, Italy. \\
$^{[c]}$ GIFT, ``Grupo de Investigaci\'{o}n en Física Te\'{o}rica", Universidad Cat\'{o}lica De La Sant\'{i}sima Concepci\'{o}n,  Av. Alonso de Ribera 2850, Concepción, Bío Bío, Chile.

\vskip 1cm

\begin{abstract}\noindent
{\small{

We propose a first-order geometric Lagrangian for four-dimensional conformal gravity within the Cartan formulation, which yields, dynamically, the standard constraints on the fields, expected for conformal gravity.
Upon imposing the dynamical constraints, together with the request of conformal invariance of the off-shell Lagrangian, the theory reduces to the standard expression  for conformal gravity, in terms of quadratic curvature invariants.
Our results clarify the geometric status of conformal gravity as a gauge theory and open the way to a similar dynamical implementation of the constraints in higher dimensions and supersymmetric extensions.

}}
\end{abstract}

\vfill
\noindent
{\footnotesize{\tt laura.andrianopoli@polito.it};\\
{\tt riccardo.dauria@polito.it}; \\
{\tt giacomo.grosso@edu.unito.it}; \\
{\tt $^*$Corresponding author, lucrezia.ravera@polito.it} }

\numberwithin{equation}{section}
\clearpage

\tableofcontents

\section{Introduction}\label{intro}

Conformal gravity is a higher-derivative theory of gravitation which is invariant under Weyl  rescalings of the metric, $g_{\mu \nu}  \rightarrow \Omega^2 \,g_{\mu \nu} $. {In four spacetime dimensions, it has a higher-derivative action  which is quadratic in the Weyl tensor \cite{Weyl:1918ib,Bach:1921zdq}.} 
Originally introduced by Weyl as an attempt to unify gravitation and electromagnetism \cite{Weyl:1918ib}, the theory later re-emerged as an alternative to Einstein gravity \cite{Stelle:1977ry}, mainly due to its improved ultraviolet behaviour, and has since been studied extensively both at the classical and the quantum levels{.
Early developments focused on the group structure and representations of conformal symmetry, 
as a possible symmetry underlying high-energy physics \cite{Barut:1971bb}, and more generally on the formulation of gauge theories for spacetime symmetries \cite{Harnad:1976yu,Neeman:1978zvv,Neeman:1978njh}. More recently, phenomenological implications of conformal (Weyl) gravity at astrophysical and cosmological scales have been studied \cite{Mannheim:2010xw,Mannheim:2011ds,Jizba:2014taa}, and the conformal symmetry has been further explored as a fundamental symmetry principle underlying spacetime dynamics \cite{tHooft:2014swy}.
The relation between conformal gravity and Einstein gravity was fruitfully analyzed: in \cite{Maldacena:2011mk} it was shown that, by the inclusion of a Neumann boundary condition, conformal gravity can be used to get the semiclassical wavefunction of the universe of four dimensional asymptotically de Sitter or Euclidean anti-de Sitter spacetimes (see also \cite{Anastasiou:2016jix}); on the other hand, the embedding of Einstein-AdS gravity in conformal gravity was shown to provide a new derivation of the Hawking mass and Willmore energy functionals for asymptotically AdS spacetimes \cite{Anastasiou:2022ljq}.
More general, non-Einstein exact solutions corresponding to black holes \cite{Liu:2012xn,Lu:2012xu}, gravitational waves \cite{Gullu:2011sj,Ayon-Beato:2012ukp} and gravitational instantons \cite{Corral:2021xsu}, were also found. 
}

The fourth-order equations of motion (e.o.m.) of Weyl's conformal gravity for the metric describe, beside  the propagation of a massless graviton of helicity $\pm 2$ and of a massless gauge vector of helicity $\pm 1$, also ghost-like states (of helicity $\pm 2$, $\pm 1$) \cite{Stelle:1976gc,Ferrara:1977mv,Ferrara:1978rk,PhysRevD.26.934,Riegert,Antoniadis:1984br}.
The standard formulation of Weyl's conformal gravity enjoys invariance under local Lorentz transformations and Weyl rescalings, but a non-manifest invariance under the full conformal symmetry is hidden. {In fact, the theory was later recognised as the gauge theory of the conformal group, where the full conformal invariance, including local special conformal transformations alongside Lorentz and dilatation symmetries, is realized at the Lagrangian level \cite{Ferrara:1977ij,Kaku:1977pa,Harnad:1976yu,Lord:1985fdw,Butter:2016qkx}.}
The resulting action, for four-dimensional conformal gravity, has a Yang--Mills-like structure. 
The same model was then expressed in a Cartan geometric framework \cite{Sharpe},
which allows a compact matrix treatment \cite{Attard:2015jfr} and a better understanding of the underlying geometry.
\\
Conformal gravity has attracted renewed interest due to its deep connections with gauge theory formulations of gravity \cite{Kaku:1977pa,MacDowell:1977jt}, its role in AdS/CFT and holography \cite{Maldacena:2011mk}, and its potential relevance to cosmology and astrophysics without invoking Dark Matter \cite{Mannheim:2005bfa}. \\
Despite these appealing features, the geometric interpretation of conformal gravity and, in particular, of its constraints, remains less developed than in the Einstein-Cartan framework for pure gravity, motivating further investigation.

This work aims   to clarify that, at least for the   $D=4$ case, it is possible to find a \emph{first-order geometric formulation}\footnote{{By ``geometric formulation" we mean a Lagrangian formulation which does not rely on a choice of the metric. This is generally obtained  if the fields are written in terms  of differential forms 
and their exterior differential ($d$), and if the kinetic terms are expressed at first-order, avoiding use of the Hodge-dual operator.}}  -- à la Cartan -- of conformal gravity, in a Lagrangian theory exhibiting  manifest invariance under the full conformal group.
By this we mean an approach where the differential and algebraic constraints usually understood \cite{Bach:1921zdq,Barut:1971bb,Harnad:1976yu,Kaku:1977pa,MacDowell:1977jt,Korzynski:2003bh,Wheeler:2013ora,Attard:2015jfr,DAuria:2021dth,Francois:2024rfh} on the fields of conformal gravity, could be obtained \emph{dynamically}, {from a \emph{Lagrangian} formulation,} just as it happens for the {torsion constraint  in the Cartan--Einstein (CE) formulation of General Relativity (GR), which arises from the spin connection equation of motion}. The above {mentioned} constraints can actually be obtained {dynamically} from a geometric Lagrangian which turns out to  be equivalent (after imposing the dynamical constraints) to the standard one of $D=4$ {conformal gravity} (CG) \cite{Kaku:1977pa,Attard:2015jfr}.
More precisely, 
besides the vielbein, the set of off-shell fields of our conformal gravity à la Cartan includes the gauge fields associated with conformal boosts and dilatations, and the Lorentz spin connection. 
Their equations of motion, which correspond to
passing to second-order, will allow to express the gauge fields in terms of the metric field and its derivatives only, thus recovering the standard Weyl theory.

The  Lagrangian that we are going to construct here   works for conformal gravity as the Cartan--Einstein Lagrangian does for gravity.
A common feature of Einstein gravity and conformal gravity is that  they are $\mathcal{H}$-gauge invariant theories based on higher $G\supset \mathcal{H}$ group structures  (the Poincaré group for Einstein gravity and the conformal group for conformal gravity), where {the gauge group $\mathcal{H}$} acts non-trivially on spacetime. In these theories, the spacetime,  whose cotangent space is spanned by the vielbein $V^a$, has actually the same dimension as the homogeneous space $\mathcal{M}= G /\mathcal{H}$, and it is locally equivalent to it.
In the vacuum, corresponding to vanishing curvatures, the identification becomes global and the full $G$-invariance of the theory is recovered.
All the above structure can be rephrased by saying that the set of gauge fields plus the vielbein are the local components of a Cartan connection \cite{CartanEspGen24,CartanProj24,CartanConf23} on a principal bundle with fiber $\mathcal{H}$ and spacetime itself as base manifold
\cite{Neeman:1978zvv,Neeman:1978njh}.\footnote{To be more precise, a Cartan bundle $P$ has \textit{structure group} $H \subset G$ spanning the fiber, and the associated gauge transformations result from the action of the bundle vertical automorphisms group ${\rm{Aut}}_v(P)$. The latter is isomorphic to the \textit{gauge group} $\mathcal{H}$. All
the relevant symmetries of theories of gravity, including the diffeomorphisms of the base manifold ${\rm{Diff}}(\mathcal{M})$, are then naturally organised in the short exact sequence $\mathcal{H} \simeq {\rm{Aut}}_v(P) \rightarrow {\rm{Aut}}(P) \rightarrow {\rm{Diff}}(\mathcal{M})$
of groups associated with the bundle. Then, in rigorous mathematical terms, the base space $\mathcal{M}$ of the bundle is locally isomorphic to the homogeneous space $G/H$.} 

Cartan geometry precisely provides the natural framework for formulating gravitational theories as gauge theories, offering a unified differential-geometric perspective on spacetime and its dynamics. 
Unlike standard Yang--Mills theories \cite{Yang:1954ek}, where an Ehresmann connection \cite{Ehresmann1950} on a principal bundle $P(\mathcal{M},H)$, with structure group $H$, defines parallel transport across the abstract fiber which is, in general, unrelated from the base manifold $\mathcal{M}$, gravity requires a connection that incorporates a \emph{soldering form}. The soldering form, which is the vielbein itself, {locally} identifies the tangent space to the base manifold with the vector space $\mathfrak{g}/\mathfrak{h}$ (where $\mathfrak{g}$ is the Lie algebra of a larger group $G\supset H$), effectively tying the fiber to the spacetime geometry itself via the vielbein 1-form $V^a ={V^a}_{\mu}dx^{\mu}$, where $x^{\mu}$ are coordinates on spacetime.
{In this framework, a crucial role is played by the \emph{Cartan connection} on the principal $H$-bundle: it is $\mathfrak{g}$-valued (with $\mathfrak{g} \supset \mathfrak{h}$), has no kernel, and induces an isomorphism $T_p\mathcal{M}\simeq \mathfrak{g}/\mathfrak{h}$. 
Its curvature splits into torsion (the $\mathfrak{g}/\mathfrak{h}$-valued part) and  $\mathfrak{h}$-curvature. 
The manifold $\mathcal{M}$ {is} locally isomorphic to the homogeneous Klein space $G/H$. 
The Klein geometry describes the (zero-curvatures) vacuum of the theory, where the full $G$-invariance is recovered as a global symmetry;  out of the vacuum, Cartan geometry provides a curved generalization of Klein geometry, well-adapted for describing gravitational dynamics. 

For the case of Einstein's theory à la Cartan, the Cartan bundle structure, if not taken \emph{a priori}, can also  be seen as emerging from the requirement of Lorentz gauge invariance and the study of the dynamics of the theory, encoded in the field equations of the gauge field.
This approach was first introduced  by Ne'eman and Regge in \cite{Neeman:1978zvv,Neeman:1978njh} from a theoretical-physics point of view, and applied to Penrose's twistor theory \cite{Penrose-McCallum72,Penrose1977,Friedrich77}, as well as to non-relativistic classical physics \cite{Duval-et-al1983} within a mathematical-physics approach.}

In this work, following the line of thought pioneered by Ne'eman and Regge \cite{Neeman:1978zvv,Neeman:1978njh} for Einstein gravity {and supergravity}, we show how the bundle structure, together with the standard constraints of conformal gravity, emerge dynamically in our theory from the field equations of a  first-order  Lagrangian. 

In this theory, the gauge group is $\mathcal{H}_{\text{C}}=(\rm{SO}(1,3)\times \rm{SO}(1,1))\ltimes \mathbb{R}^{1,3}$, and it is a non-semisimple, solvable Lie group. 
It is a subgroup of the  conformal group $G_{\text{C}}=\rm{SO}(2,4)$,  and can be obtained as an ``In\"on\"u--Wigner contraction" of it.\footnote{Note that the generators of translations  get decoupled from the rest of the algebra in the contraction. They are then projected out from $\HC$, which is a proper subgroup of $G_{\text{C}}$.} 
 The group $\mathcal{H}_{\text{C}}$ acts non-trivially on the spacetime vielbein. However, as we will see, 
the first-order Lagrangian that we will build enjoys invariance under the gauge group $\mathcal{H}_{\text{C}}$  in two different and inequivalent ways: one corresponds to thinking of $\mathcal{H}_{\text{C}}$ as embedded in the conformal group  (usually referred to as standard \emph{conformal gauge invariance}), the other to treating it as a \emph{Yang--Mills-like} gauge invariance of the Lagrangian, without referring to a larger group structure. 
 The Maurer--Cartan dual form of the algebra is different in the two cases.
The implementation of $\mathcal{H}_{\text{C}}$-invariance in this second sense is new in the study of conformal gravity, and has implications on the conformal-torsion constraint, which we are going to sketch in the following.
A key difference of conformal gravity with respect to ordinary Einstein gravity is that, in conformal gravity, the quadratic appearance of the Weyl tensor in the Lagrangian \cite{Attard:2015jfr}  makes the equation of motion for the spin connection   a genuine dynamical equation, associated with a propagating torsion, rather than an algebraic constraint. The vanishing of conformal torsion -- traditionally assumed as a kinematic constraint in the literature on conformal gravity -- cannot be obtained dynamically from the spin-connection {field equation} in the same direct way as {it happens} in the \mbox{Cartan--Einstein theory.} 

A  strong  motivation {for imposing the conformal-torsion constraint} arises from the symmetry analysis itself: the local \emph{Yang--Mills-like} invariance under $\mathcal{H}_{\text{C}}$ is not automatic {in the} first-order Lagrangian of conformal gravity, due to the non-semisimple structure of the gauge group.  
As we will see in Section  \ref{Symmetries of the Lagrangian}, the explicit calculation shows that such $\mathcal{H}_{\text{C}}$-invariance {of the first-order Lagrangian for conformal gravity} holds only when conformal torsion vanishes. 
Thus, the requirement of gauge invariance under the aforementioned structure provides a dynamical symmetry-based justification for imposing the conformal-torsion constraint.
Conversely, if invariance under this Yang--Mills–like gauge symmetry is not required, the theory may in principle admit a non-vanishing conformal torsion. 
In this case, the Yang--Mills structure of the Lagrangian is lost, whereas the Cartan bundle structure persists.

{We will however insist, here, on the symmetry argument of $\HC$-invariance of the  first-order conformal-gravity Lagrangian.  For this reason,  to implement the conformal-torsion constraint, our Lagrangian will include a two-form Lagrange multiplier that enforces vanishing {of the} conformal torsion.
As a byproduct, with the introduction of the Lagrange multiplier, then, the equations of motion of the spin connection and Weyl gauge field turn into (non-dynamical) relations expressing the Lagrange multiplier in terms of the other fields and their curvatures. }

{The analysis presented here could be fruitfully applied to more involved, higher-dimensional conformal gravity models, in particular in six dimensions, where both the group-theoretical structure and the dynamical content are substantially more intricate, and especially to build consistent supersymmetric extensions.}

{The remainder of the paper is structured as follows: 
In Section \ref{Algebraic structure}, we recall the algebraic setup of conformal gravity within the Cartan geometric framework, introducing the conformal group, its gauge subgroup, the corresponding connection and curvature components, together with the conformal Bianchi identities. 
Section \ref{The geometric Lagrangian with its dynamical constraints} presents the construction of our first-order geometric Lagrangian. 
We discuss the inclusion{, besides} of {the} auxiliary fields {of the so-called ``1.5-order formulation", to get a geometric Lagrangian which avoids the choice of a metric, also of}  a Lagrange multiplier to enforce vanishing conformal torsion, and we derive the dynamical emergence of the standard constraints from the variations of the Lagrangian with respect to the auxiliary fields, the Lagrange multiplier, and the gauge fields. 
In Section \ref{Symmetries of the Lagrangian}, we discuss the symmetries of the Lagrangian: first, its invariance under the full conformal gauge transformations within the Cartan bundle structure, showing how this requirement leads to the standard quadratic combination of curvatures; 
second, the conditions for the Yang--Mills-like invariance under the direct action of the corresponding gauge group alone. 
In Section \ref{The Lagrangian à la Cartan and its second-order expression}, we provide the final form of our Lagrangian à la Cartan and its second-order expression, 
which yields Weyl's conformal gravity.
Finally, we conclude in Section \ref{Conclusions} with a summary of the results and perspectives for higher dimensional and supersymmetric extensions.
In Appendix \ref{conf algebra}, we provide the conformal algebra and recall the Cartan kinematics, 
and, in Appendix \ref{standard conf}, we collect the standard constraints assumed in the conformal gravity literature.}

\section{Algebraic structure}\label{Algebraic structure}

The first-order Cartan formulation of conformal gravity that we are going to present here will be constructed in analogy with the Cartan formulation of Einstein gravity. To clarify similarities and differences with the case of conformal Cartan gravity, let us then first remind the relevant issues of Einstein--Cartan gravity. 

The Cartan formulation of (matter-coupled) Einstein gravity, thought of as a gauge theory whose gauge group acts non-trivially on spacetime, is based on the following specific features:
\begin{itemize}
\item
One writes a {Lorentz-invariant} Lagrangian {(built up with terms separately} invariant under the Lorentz subgroup $\mathcal{H}_{\text{E}}=\rm{SO}(1,3)$ of the Poincaré group $G_{\text{E}}= \mathrm{ISO}(1,3)$).\footnote{In general, we will denote  rigid Lorentz spacetime indices with Latin letters $a,b,\ldots=0,1,2,3$.} 
\item
{The equation of motion of the $\mathcal{H}_{\text{E}}$ gauge field, that is the condition $\frac{\delta \mathcal{L}^{\text{CE}}}{\delta\omega^{ab}}=0$, is not dynamical, enforcing  instead  the torsion constraint  which expresses the spin connection in term of the vielbein and its derivatives, thus making, in a dynamical way, the vielbein a soldering form.}

\item Furthermore,  the theory has a bundle geometric structure as a principal Cartan bundle with fiber $\mathcal{H}_{\text{E}} \subset G_{\text{E}}$ \cite{Neeman:1978zvv,Neeman:1978njh}. The base space $\mathcal{M}$ of the bundle, whose cotangent plane is spanned by the vielbein $V^a$,  is locally isomorphic to the homogeneous space $G_{\text{E}}/\mathcal{H}_{\text{E}}$ {($G_{\text{E}}$-invariance is recovered in the free-falling frame where, locally, diffeomorphisms amount to spacetime translations, that is in the zero-curvatures vacuum configuration)}.
The full set of physical 1-form fields  ($\omega^{ab}$ and $V^a$ for the case of Einstein--Cartan gravity) should then be thought of as local representatives  of the Cartan connection on the $G_{\text{E}}$-bundle:
\begin{equation}
    \mathbb{A}_{\text{E}} = {\frac{1}{2}} \omega^{ab} \mathbf{J}_{ab} + V^a \mathbf{P}_a = \Omega_{(\mathcal{H}_{\text{E}})} \oplus \mathbb{P}_{\text{E}} \,, \quad \text{taking values in } \mathfrak{g}_{\text{E}}\,,
\end{equation}
which allows the  local identification 
\begin{equation}
\Omega^{ab}_{(\mathcal{H}_{\text{E}})} =\omega^{ab}\,,\quad \mathbb{P}^a_{\text{E}}   = V^a  \,.
\end{equation}
\end{itemize}

The same recipe can be applied to conformal gravity. In this case, the  
 principal {Cartan} bundle has fiber {$\mathcal{H}_{\text{C}}  \subset G_{\text{C}}$}, where $G_{\text{C}}=\mathrm{SO}(2,4)$ is the conformal group and $\mathcal{H}_{\text{C}}=(\rm{SO}(1,3)\times \rm{SO}(1,1))\ltimes \mathbb{R}^{1,3} \subset \rm{SO}(2,4)$ 
its gauge subgroup. It is a solvable, non-semisimple ({parabolic) sub}group {of the conformal group, whose associated algebra}  can be obtained as a contraction of the conformal {algebra}, the contraction being performed on the generators with positive $\mathfrak{so}(1,1)$ grading.
The  algebra of the conformal group $\rm{SO}(2,4)$ is generated by the set of generators $\mathbf{T}_A=\lbrace \mathbf{J}_{ab}, \mathbf{P}_a, \mathbf{K}_a, \mathbf{D} \rbrace $, where we have decomposed the adjoint index $A$ of the conformal algebra with respect to the Lorentz indices $a,b,\ldots=0,1,2,3$. $\mathbf{J}_{ab}$ are the Lorentz rotations, $\mathbf{P}_a$ the spacetime translations, $\mathbf{K}_a$ the conformal boosts, and $\mathbf{D}$ the dilatations (scale transformations).
See Appendix \ref{conf algebra} for details.

An important difference of the algebraic structure of conformal gravity with respect to Einstein gravity resides in the non-semisimple nature of $\mathcal{H}_{\text{C}}$ that, as we will see, implies its two-fold possible action on the Lagrangian.
The   gauge algebra associated with $\mathcal{H}_\text{C}$    can be obtained by rescaling the translation generators, in the algebra of the conformal group $G_\text{C}$, as $\mathbf{P}_a \rightarrow \lambda \mathbf{P}_a$ with a parameter $\lambda$, then taking $\lambda \rightarrow \infty$, and reads
\begin{equation}
\label{gaugealgYM}
    \begin{split}
        & [\mathbf{J}_{cd},\mathbf{J}_{ef}] = -\frac{1}{2}(-\eta_{ce}\delta^{ab}_{fd} +\eta_{cf} \delta_{ed}^{ab} + \eta_{de} \delta^{ab}_{fc} -\eta_{df}\delta^{ab}_{ec})\mathbf{J}_{ab} = f^{[ab]}{}_{[cd] [ef]}\mathbf{J}_{[ab]} \,,\\
        & [\mathbf{J}_{bc},\mathbf{K}_d] = \frac{1}{2}(\delta^a_b \eta_{cd} -\delta^a_c \eta_{bd})\mathbf{K}_a = f^a{}_{[bc] d}\mathbf{K}_a \,, \\
        & [\mathbf{K}_a, \mathbf{D}] = \delta^b_a \mathbf{K}_b = f^b{}_{a0} \mathbf{K}_b\,.
    \end{split}
\end{equation}
{This algebra acts non-trivially on spacetime due to {fact that the generators of diffeomorphisms, which are locally isomorphic to translations, are Lorentz vectors with definite Weyl weight, implying their } non-vanishing commutators with the generators $\mathbf{J}_{ab}$ and $\mathbf{D}$.\footnote{{{Indeed the translations generator $P_a$ is a Lorentz vector, that is} $[\mathbf{J}_{bc},\mathbf{P}_d] = \frac{1}{2}(\delta^a_b \eta_{cd} -\delta^a_c \eta_{bd})\mathbf{P}_a = f^a_{\;\; [bc] d}\mathbf{P}_a$, {with a definite Weyl weight, that is} $[\mathbf{P}_a,\mathbf{D}] = -\delta_a^b \mathbf{P}_b = -f^b_{\;\; a 0} \mathbf{P}_b$.} }
Focusing on the intrinsic group-theoretical and algebraic structure of $\mathcal{H}_{\text{C}}$ independently of its embedding into the full $G_{\text{C}}$, we can then write the gauge algebra \eqref{gaugealgYM} in its dual Maurer--Cartan (MC) formulation as follows:}
\begin{subequations}\label{gauge}
\begin{align}
R^{ab} & \equiv d \omega^{ab} + {\omega^a}_c \wedge \omega^{cb} = 0 \,, 
\label{Ralg}\\
\mathring{\mathbb{G}}   & \equiv d b = 0  \,,
\label{Galg}\\
\mathbb{C}^a & \equiv \mathcal{D} S^a = dS^a + \omega^{ab} \wedge S_b - b \wedge S^a = 0 
 \,.  \label{Calg}    
\end{align}
\end{subequations}
The right-hand side of \eqref{gauge} gives the MC equations dual to the algebra \eqref{gaugealgYM}, describing the vacuum, in terms of the left-invariant 1-form fields
$\omega^{ab}$ (Lorentz spin connection),  $b$ (dilatation gauge field), and $S^a$
(special conformal 1-form field), respectively dual to the vector-field generators of the gauge algebra.\footnote{{The solvable algebra \eqref{gauge} in its dual form can be obtained as a contraction  from the MC equations of the conformal algebra, by redefining $V^a \to \frac 1\lambda V^a$, then sending $\lambda \to \infty$.}}
On the other hand, {out of the vacuum the 1-form fields become the dynamical gauge fields, whose corresponding field strengths,} on the left-hand side of \eqref{gauge}, are the Riemann tensor $R^{ab}$, the dilatation field strength $db$, and $\mathbb{C}^a$.
\\
{The above 1-form gauge fields enter the Lagrangian together with the vielbein 1-form $V^a$.}
The non-trivial action  on spacetime {of the Lorentz and Weyl subgroups of $\mathcal{H}_{\text{C}}$ manifests itself on the dynamical fields through its action on the spacetime vielbein 1-form $V^a$, that is through the conformal torsion} 
\begin{align}
    \mathbb{T}^a \equiv d V^a + {\omega^a}_b \wedge V^b + b \wedge V^a = \mathcal{D}V^a \,.
\end{align}
{However, the action of the conformal boosts of $\mathcal{H}_{\text{C}}$ on the vielbein 1-form is trivial. This can also be seen from the embedding in $G_{\text{C}}$ of the contracted group $\HC$.}

{On the other hand, if we regard $\mathcal{H}_{\text{C}}$ as embedded into $G_\text{C}$ and thus endow the theory with a principal bundle geometric structure, a (conformal) Cartan geometry naturally emerges. 
In this geometric setup, the {action of the full (non-contracted) conformal group manifests itself through the }\emph{Cartan connection, which } is described by the set of 1-form fields $\omega^{ab}$, $b$, $S^a$, and $V^a$, respectively dual to the vector-field generators of the conformal algebra, and} reads
\begin{equation}
    \mathbb{A}_{\text{C}} = {\frac{1}{2}} \omega^{ab} \mathbf{J}_{ab} + S^a \mathbf{K}_a + b \mathbf{D} + V^a \mathbf{P}_a= \Omega_{(\mathcal{H}_{\text{C}})} \oplus \mathbb{P}_{\text{C}} \quad \text{taking values in } \mathfrak{g}_{\text{C}}\,,
\end{equation}
with the local identification of
\begin{equation}
\Omega_{(\mathcal{H}_{\text{C}})} ={\frac{1}{2}} \omega^{ab} \mathbf{J}_{ab}  + S^a \mathbf{K}_a + b \mathbf{D}
\end{equation}
as the $\mathcal{H_\text{C}}$-gauge field, and of $\mathbb{P}_{\text{C}}   = V^a\mathbf{P}_a$
as the spacetime vielbein.
The field strength associated with $\mathbb{A}_\text{C}$ is
\begin{equation}
    \mathbb{F}_\text{C} ={\frac{1}{2}} \mathbb{W}^{ab} \mathbf{J}_{ab}+ \mathbb{C}^a \mathbf{K}_a + \mathbb{G} \mathbf{D} + \mathbb{T}^a \mathbf{P}_a  \,,
\end{equation}
where 
\begin{subequations}
\label{FS}\begin{align}
    \mathbb{W}^{ab} & \equiv d \omega^{ab} + {\omega^a}_c \wedge \omega^{cb} - 2 V^{[a} \wedge S^{b]} = R^{ab} - 2 V^{[a} \wedge S^{b]} = \frac 12 \,{\mathbb{W}^{ab}}_{cd} V^c \wedge V^d \,,   \label{W}\\
    \mathbb{C}^a & \equiv \mathcal{D} S^a = dS^a + \omega^{ab} \wedge S_b - b \wedge S^a =\frac 12 \, {\mathbb{C}^a}_{bc} V^b \wedge V^c \,,  \label{C} \\
    \mathbb{G} & \equiv d b -   V^a \wedge S_a = \frac 12 \,\mathbb{G}_{ab} V^a \wedge V^b \,, \label{G}\\
    \mathbb{T}^a & \equiv \mathcal{D} V^a = d V^a + {\omega^a}_b \wedge V^b + b \wedge V^a = \frac 12 \,{\mathbb{T}^a}_{bc} V^b \wedge V^c \,,   \label{T} 
\end{align}
\end{subequations}
$\mathcal{D}$ being the Lorentz plus scale covariant derivative (the Lorentz covariant derivative is instead defined as $\mathcal{D}_{\text{L}}\equiv d+\omega$). The curvature 2-forms above can be obtained through the standard formula $R^A = d\mu^A +\frac{1}{2}f^A{}_{BE}\mu^B \wedge \mu^E$, using the structure constants listed in Appendix \ref{conf algebra}.
{They describe the out-of-vacuum structure with respect to the \emph{novel} MC equations associated with this geometric structure.}
{In the dual formulation in terms of algebra generators, the extra terms in \eqref{FS}, with respect to \eqref{gauge}, correspond to the non-vanishing commutation relation in $G_\text{C}$ which are then suppressed in its contraction $\mathcal{H}_{\text{C}}$: $[\mathbf{K}_a,\mathbf{P}_b] = \eta_{ab}\mathbf{D} -2\delta^{cd}_{ab}\mathbf{J}_{cd} = (f^{KP})^0{}_{ab}\mathbf{D} +(f^{KP})^{[cd]}{}_{ab}\mathbf{J}_{[cd]}$, 
whereas within 
$\mathcal{H}_\text{C}$ one has instead 
$[\mathbf{K}_a,\mathbf{P}_b]=0$.}
This distinction will reflect in the form of the infinitesimal transformation laws of the gauge fields under these two distinct implementations of the $\mathcal{H}_\text{C}$ symmetry.

The length-scale weights {(and $\rm{O}(1,1)$-gradings)} of the 1-forms -- and of their curvatures -- are 
\begin{equation}\label{formscale}
[\omega^{ab}]=[b]=0 \,, \quad [V^a]=1 \,, \quad [S^a]=-1 \,.
\end{equation}
The Bianchi identities (BI) associated with \eqref{FS} read
\begin{subequations}
\begin{align}
    & \mathcal{D} \mathbb{W}^{ab} = -2 \mathbb{T}^{[a} \wedge S^{b]} + 2 V^{[a} \wedge \mathbb{C}^{b]} \,, \label{BianchiW} \\
    & \mathcal{D}\mathbb{T}^a = R^{ab} \wedge V_b + db \wedge V^a = \mathbb{W}^{ab} \wedge V_b + \mathbb{G} \wedge V^a \,, \label{BianchiT} \\
    & \mathcal{D}\mathbb{C}^a = R^{ab} \wedge S_b - db \wedge S^a = {\mathbb{W}}^{ab} \wedge S_b - \mathbb{G} \wedge S^a \,,  \label{BianchiC} \\
    & \mathcal{D} \mathbb{G} = {-}   \mathbb{T}^a \wedge S_a {+}   V^a \wedge \mathbb{C}_a \,.
    \label{BianchiG}
\end{align}
\end{subequations}

Then, concerning the dynamics we are going to consider, the constraints required on the Lagrangian $\mathcal{L}^{\text{CG}}$ of conformal gravity (that is, {constraining on-shell the gauge fields such that} $S^a_\mu = S^a_\mu( V^a, \partial V^a)$, $\omega^{ab}= \omega^{ab}( V^a, \partial V^a, \mathbb{T}^a{}_{bc})$, $b=0$), should be obtained dynamically, from 
$\frac{\delta \mathcal{L}^{\text{CG}}}{\delta\omega^{ab}}=0$, $\frac{\delta \mathcal{L}^{\text{CG}}}{\delta b}=0$, $\frac{\delta \mathcal{L}^{\text{CG}}}{\delta S^{a}}=0$.
Upon imposing the dynamical constraints, the Lagrangian $\mathcal{L}^{\text{CG}}$ should reduce to the standard one of conformal gravity. 
We collect in Appendix \ref{standard conf} all the constraints that are assumed to hold kinematically in the standard approach.

\section{The geometric Lagrangian with its dynamical constraints}\label{The geometric Lagrangian with its dynamical constraints}

To construct a Lagrangian capable of implementing the conformal gravity constraints dynamically, we  should write, to start with, the most general sum of Lorentz and scale invariant 4-form terms, with the same parity of the Einstein--Cartan Lagrangian, refraining from assuming off-shell the  validity of the standard constraints. 

Furthermore, for later convenience, we will formulate our Lagrangian as a geometric Lagrangian, which shall not depend on a specific reference frame. To this aim, we work at first-order for the field strengths, so as to avoid use of the Hodge-duality operator in the Lagrangian.
To be fully general, we also allow for boundary terms to play a role.

Let us then write the following linear combination of Lorentz and scale invariant 4-form terms,  written in terms of the field strengths \eqref{FS}, a linear combination of which should give our locally $\mathcal{H}_{\text{C}}$-invariant Lagrangian:
\begin{align}
\label{LCG}  
\mathcal{L}^{\text{CG}} =& a_1  \mathbb{W}^{ab} \wedge   \mathbb{W}^{cd} \epsilon_{abcd}+
   a_2 \mathbb{W}^{ab} \wedge  \widetilde{W}_{ab}{}^{|cd} V^\ell \wedge V^m \epsilon_{cd\ell m} -\frac 1{24} a_2 
 \widetilde{W}^{ab}{}_{|cd}\widetilde{W}_{ab}{}^{|cd}\Omega^{(4)}
\nonumber\\
&  +b_1\mathbb{W}^{ab} \wedge   V^c\wedge S^d \epsilon_{abcd}+
b_4 \,V^a\wedge V^b\wedge S^c\wedge S^d \epsilon_{abcd}\nonumber\\
&+
c_1\left(   \mathbb{T}^a\widetilde{C}_a{}^{\ell m} +  \mathbb{C}^a\widetilde{T}_a{}^{\ell m}\right)\wedge V^c\wedge V^d \epsilon_{\ell m cd} - \frac{1}{12}  c_1 \widetilde{T}_a{}^{bc}\widetilde{C}^a{}_{bc}\Omega^{(4)}\nonumber\\
&+
d_1  \mathbb{G}\, \widetilde{G}^{ab}\wedge V^c\wedge V^d \epsilon_{abcd} -\frac 1{24}  d_1 \widetilde{G}^{ab}\widetilde{G}_{ab}\Omega^{(4)}+
V^a \wedge \mathcal{D}\Phi_a\,,
\end{align}
where $\Omega^{(4)}\equiv V^{\ell_1}\wedge \cdots \wedge V^{\ell_4}\epsilon_{\ell_1\cdots \ell_4}= -24\, \sqrt{-g}d^4x$.\footnote{{Let us remark that, in general, $\mathbb{W}_{cd}  \epsilon^{abcd}\neq {}^*\mathbb{W}^{ab}$. 
Indeed, decomposing them on a basis of 2-forms $ V^\ell \wedge V^m\equiv \epsilon^{ab\ell m}\Omega^{(2)}_{ab}$, and using for the Hodge-duality:
$   {}^* (V^{i_1}\wedge \dots \wedge V^{i_k})=\frac{1}{(D-k)!}\,V^{j_1}\wedge \dots \wedge V^{j_{D-k}}\,\epsilon_{j_1\dots j_{D-k}}{}^{i_1\dots i_k}$ (satisfying, in $D=4$,  ${}^*{}^*=-1$), 
we have
\begin{align*}
\mathbb{W}^{ab}  \epsilon_{abcd}&=
\frac 12 \mathbb{W}^{ab}{}_{\ell m}\epsilon_{abcd} \epsilon^{\ell m rs}\Omega^{(2)}_{rs}=-2\left(\mathbb{W}^{ab}{}_{cd}\Omega^{(2)}_{ab} -4 \mathbb{W}^{ab}{}_{b[c}\Omega^{(2)}_{ d]a}+\mathbb{W}^{ab}{}_{ab}\Omega^{(2)}_{cd}
\right)\,,\\
^*\mathbb{W}_{cd}&=\frac 14 \mathbb{W}_{cd}{}^{ab}\epsilon_{ab\ell m }\epsilon^{\ell m rs} \Omega^{(2)}_{rs}= -\mathbb{W}_{cd}{}^{ab}\Omega^{(2)}_{ab}\,.
\end{align*}
In the standard formulation of conformal gravity, where $\mathbb{W}^{ab}{}_{cd}=\mathbb{W}_{cd}{}^{ab}$ and $\mathbb{W}^{ab}{}_{cb}=\mathbb{W}^{ab}{}_{ab}=0$, the above relations boil down to $\mathbb{W}^{ab}  \epsilon_{abcd}= 2  ^*\mathbb{W}_{cd}$.
}}
The Lagrangian \eqref{LCG} depends on the spacetime vierbein $V^a$, the gauge 1-form fields $\omega^{ab}, b, S^a$, and on the auxiliary tensorial 0-form fields $\widetilde{W}^{ab}{}_{cd}$, $\widetilde{T}^a{}_{bc}$, $\widetilde{C}^a{}_{bc}$, $\widetilde{G}_{ab}$, which allow to write the Lagrangian geometrically.\footnote{{The coefficients, in front of the terms quadratic in the auxiliary tensorial 0-forms, set the normalization of the latter, in such a way  that the auxiliary fields get exactly identified, through their field equations, with the spacetime components of the corresponding  physical fields (see eqs. \eqref{aux}).}}

\paragraph{Vanishing conformal torsion\\}

Besides the above mentioned fields, our Lagrangian also depends on the 2-form Lagrange multiplier $\Phi^a\equiv \frac 12 \Phi^a{}_{bc}V^b\wedge V^c$, implementing the condition $\mathbb{T}^a=0$.
This is a sort of ``compromise", in order to reproduce the conformal-torsion constraint, the latter being one of the standard constraints which are  generally assumed to hold in conformal gravity.
Indeed, differently from the case of the Cartan--Einstein theory, where the vanishing torsion constraint directly follows from the spin connection e.o.m., the Weyl tensor (which is the field strength of the spin connection $\omega^{ab}$) appears quadratically in the  Lagrangian of conformal gravity, making in this case the spin connection e.o.m.  a dynamical equation, and not a constraint.

An {algebraic} motivation to impose the conformal-torsion constraint emerges from our analysis. Indeed, as we will see in detail  in Section \ref{Checking conformal invariance}, {the vanishing of the conformal torsion} is a necessary condition in order for our Lagrangian to be $\mathcal{H}_\text{C}$-invariant.
The condition of $\mathcal{H}_{\text{E}}$ (Lorentz)-invariance is automatically enjoyed by the Cartan--Einstein Lagrangian of matter-coupled gravity theories, which indeed is built up of Lorentz-invariant terms. On the other hand, in the case of  conformal gravity, where the  $\mathcal{H}_\text{C}$ gauge group is \emph{non-semisimple}, $\mathcal{H}_\text{C}$-invariance of our first-order conformal gravity  Lagrangian \eqref{LCG} is not obviously guaranteed, 
and it is indeed satisfied only after imposing the vanishing of $\mathbb{T}^a$.\footnote{We notice that, on the other hand, $G_{\text{C}}$ is simple, while $G_{\text{E}}$ is not.} 

{On the other hand, if we would not impose the torsion constraint via a Lagrange multiplier, thus keeping  a possible contribution to the field equations from the conformal torsion, still a careful analysis of the gauge fields e.o.m. shows that none of the possible irreducible components of the conformal-torsion tensor is allowed in our first-order Lagrangian.}

\subsection{Variation with respect to the auxiliary fields}

{By varying our first-order Lagrangian \eqref{LCG} with respect to the auxiliary fields, we get}
\begin{subequations}\label{aux}
\begin{align}
\frac{\delta \mathcal{L}^{\text{CG}}}{\delta \widetilde{W}^{ab}{}_{cd}}=0 \,:&\quad \widetilde{W}^{ab}{}_{cd}=\mathbb{W}^{ab}{}_{cd}
\,,\label{auxw}\\
\frac{\delta \mathcal{L}^{\text{CG}}}{\delta \widetilde{G}_{ab}}=0 \,:&\quad   \widetilde{G}_{ab}= \mathbb{G}_{ab} 
\,,\label{auxg}\\
\frac{\delta \mathcal{L}^{\text{CG}}}{\delta \widetilde{T}^a{}_{bc}}=0 \,:&\quad \widetilde{C}^a{}_{bc}= \mathbb{C}^a{}_{bc} \,,\\
\frac{\delta \mathcal{L}^{\text{CG}}}{\delta \widetilde{C}^a{}_{bc}}=0 \,:&\quad \widetilde{T}^a{}_{bc}= \mathbb{T}^a{}_{bc}\,.
\end{align}
\end{subequations}
On the other hand, variation with respect to the Lagrange multiplier,
\begin{align}
\frac{\delta \mathcal{L}^{\text{CG}}}{\delta \Phi^a}=0 \,:&\quad \mathbb{T}^a{}= 0 
\,.\label{tor0}
\end{align}
Note that the condition \eqref{tor0} also implies, for consistency, $\mathcal{D}\mathbb{T}^a=0$, that is
\begin{align}
  \mathbb{W}^{ab}\wedge V_b + \mathbb{G} \wedge V^a=0\,,  
\end{align}
which further implies 
\begin{align}
  \mathbb{W}^{b}{}_{[c|d]b}= - \mathcal{R}_{[cd]}+ 2 S_{[c|d]} = -\mathbb{G}_{cd} \quad \Leftrightarrow \quad \mathcal{R}_{[cd]} = 2 {\mathcal{D}_{[c}}  b_{d]} \,. \label{anti}
\end{align}

\subsection{Variation with respect to the gauge fields}

Let us now analyze the field equations of the dynamical fields. As we are going to see, having included in the Lagrangian \eqref{LCG} the term with the Lagrange multiplier $\Phi^a$, the only relevant conditions come from the $S^a$ gauge field, the other field equations just amounting to express $\Phi^a$ in terms of the physical fields. 
We get the following results.
\begin{itemize}
\item \label{ds} 
$\frac{\delta \mathcal{L}^{\text{CG}}}{\delta S^a}=0$: From the variation of the Lagrangian with respect to $S^a$ we obtain
\begin{align}
& (b_1-4a_1)V^b \wedge \mathbb{W}^{cd} \epsilon_{abcd} -2 a_2 V^b \wedge \widetilde{W}_{ab}{}^{cd} V^\ell\wedge V^m \epsilon_{cd \ell m} \nonumber\\ &- 2(b_1-b_4) S^b\wedge V^c\wedge V^d\epsilon_{abcd}+c_1\mathcal{D}(\widetilde{T}_a{}^{\ell m}V^c\wedge V^d\epsilon_{cd\ell m})\nonumber\\
&+d_1\widetilde{G}^{\ell m}
V_a\wedge V^c\wedge V^d\epsilon_{cd\ell m}=0 \,,
\end{align}
that is, in components and using \eqref{tor0}, 
\begin{align}
  &  (b_1-4a_1)\mathbb{W}^{cd}{}_{|cd}\eta_{at}
    -(8a_1  -2b_1)\mathbb{W}^{b}{}_{t|ab} + \,8 a_2\widetilde{W}^{b}{}_{a|tb} \nonumber\\
    &- 4 d_1(\tilde{G}_{at})    + 4(b_1-b_4)\left( S_{t|a}-S^b{}_{|b} \eta_{at}\right)=0
\,.   \label{seq}
\end{align}
As a first observation, let us remark that  {a particular solution to the field equations has to be the zero-curvatures vacuum.} In this case, eq. \eqref{seq} reduces to
\begin{align}
    4(b_1-b_4)\left( S_{t|a}-S^b{}_{|b} \eta_{at}\right)=0 \,,
\end{align}
which implies  {the following condition on the coefficients  of the Lagrangian \eqref{LCG}}:
\begin{align}
b_1=b_4\,.
\label{b14}
\end{align}

Eq. \eqref{seq} can then be decomposed in its symmetric and antisymmetric parts in the indices $a,t$, giving, after substituting \eqref{aux},
\begin{align}
(at):\,\, &   (b_1-4a_1)\mathbb{W}^{cd}{}_{|cd}\eta_{at}
    -(8a_1 - 8a_2 -2b_1 )\mathbb{W}^{b}{}_{(t|a)b} =0
    \,, \label{ss}\\
[at]:\,\,  & \, (2b_1-8a_1 - 8a_2 )\mathbb{W}^{b}{}_{[t|a]b} 
+ 4 d_1\mathbb{G}_{ta}
=0 \,.  \label{sas}
\end{align}
Comparison of eq. \eqref{sas} with  eq. \eqref{anti}, at $\mathbb{T}^a=0$ implies the condition
\begin{align}
  & 2d_1= b_1-4a_1 {- }\, 4a_2\,.\label{d1} 
\end{align} 
On the other hand,   
 taking the trace of eq. \eqref{ss} gives
\begin{align}
    2 \left(b_1-4a_1-4 a_2 \right)\mathbb{W}^{cd}{}_{|cd} = 0 \,.
\end{align}
For $b_1\neq 4(a_1+a_2) $, 
this implies $\mathbb{W}^{cd}{}_{|cd}=0$ and then, if further $b_1\neq 4(a_1-a_2)$, also \begin{align}\mathbb{W}^{b}{}_{(t|a)b}=0\,.
\end{align}
This is one of the standard constraints that hold for conformal gravity  at $\mathbb{T}^a=0$, since
it implies
\begin{align}
    \label{schouten}S_{(a|b)}=\frac 1{2} \left(\mathcal{R}_{(ab)}-\frac 1{6} \eta_{ab}\mathcal{R}\right)\,,\quad S^a{}_{|a}\equiv S = \frac 16 \mathcal{R} \,,
\end{align}
which, in Weyl's conformal gravity, defines $S_{(a|b)}$ as the Schouten tensor.
On the other hand, regarding the curvature component ${\mathbb{W}^c}_{[a|b]c} = - \mathcal{R}_{[ab]} + 2S_{[a|b]}$, 
which at $\mathbb{T}^a=0$ satisfies:
\begin{align}  {\mathbb{W}^c}_{[a|b]c} = -\mathbb{G}_{ab}= - \mathcal{R}_{[ab]} + 2S_{[a|b]}\,,
    \end{align}we do not achieve the other constraint usually assumed in conformal gravity, that is 
    $\mathbb{W}^{c}{}_{[a|b]c}=0$, which would imply
  $ S_{[a|b]}=\frac 1{2} \mathcal{R}_{[ab]}$.

  \paragraph{Making the Weyl symmetry global (first part)\\} 

As it is well known \cite{Callan:1970ze,Coleman:1970je,Polchinski:1987dy,Nakayama:2013is,Fioresi:2015bta}, when conformal gravity is expressed solely in terms of the Weyl tensor as a higher-derivative theory for the metric field, the gauge symmetry associated with Weyl rescalings  is turned into a global symmetry.

As we are going to show, this feature
can be understood in our formalism, at $\mathbb{T}^a=0$,   by comparing the field equation of the gauge field $S^a$   with    the set of Bianchi identities of the model.
 As we shall discuss in detail in Section \ref{conformal_invariance}, this issue will be further clarified by passing to the second-order expression for the  Lagrangian at $\mathbb{T}^a=0$,    upon imposing the equations of motion of the gauge fields. 

Indeed, let us recall that the Bianchi identity for  the Lorentz curvature implies that the Einstein tensor
\begin{align}
\mathcal{G}_{ab}\equiv    \mathcal{R}_{ab} - \frac 12 \eta_{ab}\mathcal{R}
\,,
\end{align}
whose symmetric part is
\begin{align}
\mathcal{G}_{(ab)}=2\left(S_{(a|b)}- \eta_{ab}S^c{}_{|c}\right)    \,,
\end{align}
satisfies  the following condition:
\begin{align}\mathcal{D}_L^a\mathcal{G}_{ab}=0\,.\label{ein}\end{align}This condition follows from the Bianchi identity of the Riemann tensor, $\mathcal{D}_L R^{ab}=0$, and it holds true even when the Ricci tensor has an antisymmetric part.\footnote{In terms of $S_{a|b}$, in $D$ spacetime dimensions it gives: $\mathcal{G}_{(ab)}=(D-2)\left(S_{(a|b)}-\eta_{ab}S^c{}_{|c}\right)$.} 
We now observe that 
this condition corresponds to turning the Weyl symmetry into a global symmetry. 
Indeed,
\begin{align}
    \mathbb{W}^{b}{}_{a|cb}=\mathbb{W}^{b}{}_{[a|c]b} \,,
\end{align}
and, at $\mathbb{T}^a=0$, from the BI we get 
\begin{align}
 \label{relb}   \mathcal{D}^a \mathbb{W}^{b}{}_{a|cb}= 2 \mathbb{C}^a{}_{|ac}\,.
\end{align}
Then, using the definitions \eqref{W} and \eqref{Calg}, we get
\begin{align}
{\mathbb{W}^b}_{a|cb}=&\,{\mathbb{W}^b}_{[a|c]b}   = 2 S_{a|c}+\eta_{ac}S^b{}_{|b}- \mathcal{R}_{ac} \,, \\
  {\mathbb{C}^a}_{a|c}  =& \,\mathcal{D}_a\left(S^a{}_{|c}-\eta_{ac} S^b{}_{|b} \right) \,,
\end{align}
and, substituting in \eqref{relb},
 we obtain
 \begin{align}
     \mathcal{D}^a\left(\mathcal{R}_{ac}-\frac 12 \eta_{ac} \mathcal{R}\right)=0\,,
 \end{align}
that is  
\begin{align} \mathcal{D} ^a  \mathcal{G}_{ac}=0\,.\label{cru}
\end{align}
Notice that it is compatible with eq. \eqref{ein} only  if the Weyl connection $b$ is not minimally coupled with $\mathcal{G}_{ab}$, despite of the fact that it carries a non-vanishing Weyl weight. 

This is an indication that the Weyl symmetry, when acting on the second-order fields (that is expressed in terms of the metric and its derivatives), becomes a global symmetry. 
This argument will be further clarified when inspecting the final form of the Lagrangian, written at second-order for the gauge fields, at $\mathbb{T}^a=0$, in Section \ref{conformal_invariance}. 

\item \label{db}
$\frac{\delta \mathcal{L}^{\text{CG}}}{\delta b}=0$: The variation of the Lagrangian with respect to the gauge field $b$ yields
\begin{align}
    c_1 ( V^a \widetilde{C}_a{}^{\ell m} - S^a \widetilde{T}_a{}^{\ell m}) V^c V^d \epsilon_{\ell m cd}+\,d_1 \mathcal{D}(\widetilde{G}^{ab}V^c V^d )\epsilon_{abcd}
    + V^a\wedge \Phi_a=0 \,,
\end{align}
that is, using the constraints \eqref{aux} and \eqref{tor0},
\begin{align}\label{tra2}
    4c_1  \mathbb{C}_a{}^{at}+
   4d_1 \mathcal{D}_a  \mathbb{G}^{at}
    = \tilde\Phi_a{}^{at} \,,
\end{align}
where we have defined the Hodge-dual of the Lagrange multiplier 2-form as
\begin{align}
\label{phitphirel}
    \tilde \Phi^a{}_{|bc} \equiv\frac 12 \Phi^{a|\ell m}\epsilon_{\ell m bc}  \,.
\end{align}
Using now the conditions from the Bianchi identities at 
$\mathbb{T}^a=0$,  
then, eq. \eqref{tra2} gives
\begin{align}\label{traphi}
\tilde\Phi_a{}^{at}=4(c_1{-}2d_1)C_a{}^{at}\,.
\end{align} 

\item \label{dw}
$\frac{\delta \mathcal{L}^{\text{CG}}}{\delta \omega^{ab}{}}=0$: From the variation of the Lagrangian with respect to the spin connection
\begin{align}
    & { (b_1-4a_1)}(\mathbb{T}^c S^d-V^c \mathbb{C}^d)\epsilon_{abcd}
    +a_2(\mathcal{D}\widetilde{W}_{ab}{}^{cd}  V^\ell +2\widetilde{W}_{ab}{}^{cd} \mathbb{T}^\ell)\wedge V^m \epsilon_{cd\ell m} 
    \nonumber\\
    &+ c_1( e_{[b} \widetilde{C}_{a]}{}^{\ell m}+S_{[b} \widetilde{T}_{a]}{}^{\ell m})\wedge V^c\wedge V^d \epsilon_{cd\ell m}  -e_{[a}\wedge\Phi_{b]} =0\,.
\end{align}
Using the constraints \eqref{aux}, \eqref{tor0}, it gives
 {\begin{align}
    { (b_1-4a_1)}(\mathbb{C}^\ell{}_{|ab}+ 2\mathbb{C}^c{}_{|c[a}\delta^\ell_{b]}) {- 4a_2 \mathcal{D}_c \mathbb{W}_{ab}{}^{c\ell}}-4 c_1 \mathbb{C}_{[a|b]}{}^\ell
+\frac 12 \Phi^{[a}{}_{cd}\epsilon^{b]cd\ell}=0\,,
\end{align}
that is, recalling  \eqref{phitphirel},
\begin{align}\label{phianti}
\tilde\Phi_{[a|b]}{}^{\ell}={ (4a_1-b_1)}(\mathbb{C}^\ell{}_{|ab}+ 2\mathbb{C}^c{}_{|c[a}\delta^\ell_{b]}) {+ 4a_2 \mathcal{D}_c \mathbb{W}_{ab}{}^{c\ell}}+4 c_1 \mathbb{C}_{[a|b]}{}^\ell
\,.    
\end{align}
Let us take its trace, to be compared with \eqref{traphi}.
}
\\
Using $\mathbb{T}^a=0$ and ${\mathbb{W}^{a}}_{(b|c)a}=0$, we get
\begin{align}
  \tilde{\Phi}^a_{\,\,\,at} = - 4 [-(4a_1-b_1+c_1){\mathbb{C}^a}_{|at}- 2 a_2 \mathcal{D}^a {\mathbb{W}^c}_{[a|t]c}] \,.
\end{align}
Since, at $\mathbb{T}^a=0$, the BI of $\mathbb{G}$ and \eqref{antiw} imply  
\begin{align}
\label{BIconstr}
    2 {\mathbb{C}^a}_{ab}=\mathcal{D}^a {\mathbb{W}^c}_{[a|b]c}=-\mathcal{D}^a \mathbb{G}_{ab}\,,
\end{align}
this can be rewritten as
\begin{align}
\label{Phitdeltaom}
    \tilde{\Phi}_a^{\,\,\,at} = 4 (4a_1 - b_1 + c_1 + 4 a_2){\mathbb{C}^a}_{|at} \,.
\end{align}
 Comparison with \eqref{traphi} gives the condition
\begin{align}
 2d_1= b_1 -4a_1   - 4 a_2\,.   
\end{align}
On the other hand, directly from \eqref{phianti}, we get
\begin{align}  \label{inu} \tilde{\Phi}^a_{\,\,\,bc}  = & (b_1-4a_1+4c_1) {\mathbb{C}^a}_{|bc} + 2 (4a_1-b_1) \left({\mathbb{C}_{[b|c]}}^a-2 {\mathbb{C}^d}_{|d[b}\delta_{c]}^a\right) \nonumber\\    &+4 a_2 \mathcal{D}_d (- 2\mathbb{W}^a{}_{[b|c]}{}^d + {\mathbb{W}_{bc}}^{|ad} )\,,
\end{align}
whose trace reproduces \eqref{traphi}.

\end{itemize}

\subsection{Summary of our results}

The results obtained dynamically so far from the Lagrangian \eqref{LCG}, which include the constraint $\mathbb{T}^a=0$, are the following.
\begin{itemize}
    \item From the condition $\mathbb{T}^a=0$ we get:
    \begin{align}
  \mathbb{W}_{[ab|cd]}=0\,,\quad \mathbb{W}^{a}{}_{[b|c]a}=-\mathbb{G}_{bc}\,, \quad \mathbb{C}^a{}_{|ab}=\frac 12 \mathcal{D}^a \mathbb{W}^c{}_{a|bc}   \,, \quad \mathcal{R}_{[ab]} = 2 \mathcal{D}_{[a} b_{b]}  \,; \end{align}
  \item The conditions $\mathbb{W}^{a}{}_{(b|c)a}= 0$ and $\mathbb{W}^{a}{}_{[b|c]a}=-\mathbb{G}_{bc}$ imply:
  \begin{align}
      & S_{(a|b)}=\frac 1{2} \left(\mathcal{R}_{(ab)}-\frac 1{6} \eta_{ab}\mathcal{R}\right)\,,\\
      & S_{[a|b]} = \frac 12\left(\mathcal{R}_{[ab]}-\mathbb{G}_{ab}\right) 
      \,; \label{antiw}
  \end{align}
  \item The Lagrange multiplier 2-form is expressed in terms of the field strengths of $S^a$ and $\omega^{ab}$ as in \eqref{inu};
  \item Finally, the above conditions require the following relations among the coefficients of the Lagrangian \eqref{LCG}:
  \begin{align}    b_4=b_1\,,\quad b_1-4a_1-4a_2= 2d_1 \,,\quad b_1\neq 4(a_1 + a_2)\,, \quad b_1 \neq 4(a_1-a_2) \,.\label{coe}
  \end{align}
\end{itemize}

Let us remark that in the ``standard" formulation of conformal gravity, besides $\mathbb{T}^a=0$, one also assumes $\mathbb{W}^{ab}{}_{|cb}=0$, $S_{[a|b]}=0$, $\mathbb{G}_{ab}=0$ and $\mathcal{R}_{[ab]}=0$, from which it follows $\mathbb{C}^a{}_{|ab}=0$, $\mathbb{C}_{[a|bc]}=0$.
 Within the present approach, we are not able to get the vanishing of these  quantities, at first-order. However, they are all related to each other, since
 \begin{equation}\label{tru}
  \mathbb{W}^a{}_{b|ca}=  \mathbb{W}^a{}_{[b|c]a} =- \mathbb{G}_{bc}= 2S_{[b|c]} - \mathcal{R}_{[bc]} \,,
 \end{equation}
and
\begin{align}
\label{c00}\mathbb{C}^a{}_{ab}=\frac 12 \mathcal{D}^a \mathbb{W}^c{}_{a|bc}\,.
\end{align}
 As we will see below, in Section \ref{conformal_invariance}, however, for $\mathbb{T}^a=0$ this sector gets decoupled from the fields of the standard theory, and can be truncated out from the theory.
 Furthermore, at $\mathbb{T}^a=0$, once the Lagrangian is expressed  at second-order for the gauge fields, the sector involving
 the fields in \eqref{tru} completely disappears from the Lagrangian. 
For comparison, let us mention the first-order formulation of conformal gravity proposed in \cite{Dneprov:2022jyn}, where the theory is written in terms of a conformal-algebra-valued connection supplemented by auxiliary fields carrying the index structure of the Weyl and Cotton tensors. 
In that approach, the algebraic properties of the auxiliary sector ensure that the dilation and special conformal components drop out, while the remaining constraints on torsion and the Lorentz connection are recovered through field equations and algebraic gauge symmetries, leading to the Weyl-squared action after gauge-fixing. 
The primary focus there is on exhibiting the presymplectic BV-AKSZ structure of the model, which provides directly the corresponding BV formulation within a geometric framework (see also \cite{Dneprov:2024cvt} for a more refined exposition of the presymplectic BV-AKSZ approach in  this context). \\
In our first-order conformal gravity theory, as far as the Lagrange multiplier $\Phi^a$ is concerned, in the $\mathbb{T}^a=0$ case its on-shell expression,  eq. \eqref{inu},  simplifies into
\begin{align}
\label{PhitexprC}
{\Phi}^a 
= - 4 (d_1+ c_1){}^* \mathbb{C}_a \,,
\end{align}
where 
\begin{align}
   {}^* \mathbb{C}_a = \frac{1}{4} {\mathbb{C}^a}^{\ell m} V^b V^c \epsilon_{bc \ell m} \,,
\end{align}
and we have used \eqref{coe}.

 Using the relations found for the coefficients, we can   rewrite the Lagrangian \eqref{LCG}, upon using \eqref{aux}, but not yet \eqref{tor0}, as
\begin{align}
\label{LCG3}  
\mathcal{L}^{\text{CG}} =& a_1  \left(\mathbb{W}^{ab}+2V^a\wedge S^b\right) \wedge  \left( \mathbb{W}^{cd}+2V^c\wedge S^d\right) \epsilon_{abcd} +
  2\, a_2 \mathbb{W}^{ab} \wedge  ^*\mathbb{W}_{ab}  \nonumber\\
&
+ (2d_1+4a_2)\left(\mathbb{W}^{ab}+ \,V^a\wedge S^b\right) \wedge   V^c\wedge S^d  \epsilon_{abcd}+
4 c_1  \mathbb{T}^a\wedge ^* \mathbb{C}_a +
2d_1 \mathbb{G} \wedge ^*\mathbb{G}
\,,
\end{align}
that we supplement with the Lagrange multiplier term $V^a\mathcal{D}\Phi_a$ to implement the torsion constraint.
Note that the first term in \eqref{LCG3} is the Euler boundary term.
The remaining  free coefficients shall now be  further restricted by the requirement of invariance under conformal boosts.

\section{Symmetries of the Lagrangian}\label{Symmetries of the Lagrangian}

We shall now study the symmetries of the Lagrangian \eqref{LCG3}{, whose algebraic structure was the object of Section \ref{Algebraic structure}}. 
By construction, each term of the Lagrangian is automatically invariant under coordinate transformations and under Lorentz and Weyl gauge transformations. On the other hand, invariance under  conformal boosts is not automatic, and  must  be required explicitly to  get full conformal invariance.

We first focus on \emph{conformal invariance}, meaning invariance of the Lagrangian under the gauge subgroup $\mathcal{H}_\text{C}$ {embedded in} the conformal group as a principal Cartan bundle.
For more details on the structure of the conformal group and, in particular, its (structure group and) conformal gauge subgroup, and the conformal Cartan kinematics, we refer the reader to Appendix \ref{appconf}. Here we will consider infinitesimal transformations.

Then, in the second part of this section, {we will} inspect under which conditions our Lagrangian can be invariant also under the direct action of the gauge group $\mathcal{H}_{\text{C}}$, without referring to the underlying Cartan bundle structure.  
As clarified in Section \ref{Algebraic structure}, this corresponds to performing a contraction on the generators of translations in the conformal group.

\subsection{Conformal gauge invariance}\label{conformal_invariance}

Under special conformal transformations \cite{Ogiue,Kobayashi1972,Sharpe,capslovak:2009}, with parameter $k^a$, $k$ for short, the fields of  Cartan-conformal gravity transform as
\begin{align}
\label{gaugetrconf}
    & \delta_k \omega^{ab} = - 2 V^{[a} k^{b]} \,, \nonumber \\
    & \delta_k V^a = 0 \,, \nonumber \\
    & \delta_k S^a = \mathcal{D} k^a \,, \nonumber \\
    & \delta_k b = -  V^a k_a \,.
\end{align}
From \eqref{gaugetrconf} we also find the transformations of their field strengths:
\begin{align}
    & \delta_k \mathbb{W}^{ab} = - 2 \mathcal{D} V^{[a} k^{b]} \,, \nonumber \\
    & \delta_k \mathcal{D} V^a = 0 \,, \nonumber \\
    & \delta_k \mathcal{D} S^a = \mathbb{W}^{ab} k_b - \mathbb{G} k^a \,, \nonumber \\
    & \delta_k \mathbb{G} = {-}  \mathcal{D} V^a k_a \,.
\end{align}
We shall now use these transformations to check under which conditions  the Lagrangian \eqref{LCG3} is invariant. 
We get
\begin{align}
  \delta_k \mathcal{L}^{\text{CG}} =& k_a \mathbb{T}_b \wedge \left({-}(4a_2+2d_1)\mathbb{W}_{cd}  \epsilon^{abcd}+4(2a_2-c_1)
 ^*\mathbb{W}^{ab}\right)
-4 (c_1+d_1)k^a \mathbb{G} \wedge ^*\mathbb{T}_a \nonumber\\
&-(4a_2 + 4 a_1  + 2d_1)d\left[V^ak^b\left(\mathbb{W}^{cd}+2V^c\wedge S^d\right)  \epsilon_{abcd}\right]
\,.\label{deltaL}
\end{align}
Special conformal invariance  of $\mathcal{L}^{\text{CG}}$, up to boundary terms, would thus require, regardless of whether or not the conformal torsion vanishes: 
    \begin{align}
    \label{confinvnulkcoeff}
        c_1 = 2a_2 =- d_1   \,.
    \end{align}

Let us notice that the inclusion in the Lagrangian of the Lagrange multiplier term $V^a\mathcal{D}\Phi_a$, enforcing the torsion constraint, should not alter the conformal invariance of the Lagrangian itself. Since $\delta_k \mathbb{T}^a=0$, this implies that the multiplier $\Phi^a$ should be invariant under conformal boosts. Remarkably, this agrees with the on-shell expression \eqref{PhitexprC} found for it at $\mathbb{T}^a=0$, using \eqref{confinvnulkcoeff}.

\subsection{Gauge invariance à la Yang--Mills}\label{Checking conformal invariance}

The Cartan--Einstein Lagrangian is built up with $\mathcal{H}_\text{E}$-invariant terms. On the other hand, the invariance of the Cartan conformal gravity under the  gauge group $\mathcal{H}_{\text{C}}$ is not automatically satisfied, because of its  non-semisimple structure.

Therefore, we now study the conditions for the invariance of our conformal gravity theory à la Cartan under the gauge group $\mathcal{H}_{\text{C}}= (\rm{SO}(1,3) \times \rm{SO}(1,1))\ltimes \mathbb{R}^{1,3}$ without referring to the underlying Cartan $G_{\text{C}}$-bundle structure.

The gauge algebra we now consider is \eqref{gaugealgYM}, and the associated field strengths are \eqref{gauge}, which satisfy the following Bianchi identities:
\begin{subequations}\label{gaugebi}
\begin{align}
\mathcal{D}_L R^{ab} & =0\,,   \label{dRalg}\\
d\mathring{\mathbb{G}}   & =0 \,, \label{dGalg}\\
 \mathcal{D}  \mathbb{C}^a & =R^{ab} \wedge S_b - \mathring{\mathbb{G}} \wedge S^a \,,  \label{dCalg}    
\end{align}   
\end{subequations}
where  $\mathcal{D}_L$ denotes the Lorentz-covariant derivative.

The infinitesimal transformation of the physical fields under the transformations generated by the tangent vector dual to $S^a$ above, with parameter $\tilde\kappa$, read\footnote{Notice that these are not the special conformal transformations, as we can see by comparison with Section \ref{conformal_invariance}.}
\begin{align}
\label{altertrasf}
  & \delta_{\tilde\kappa} S^a = \mathcal{D} {\tilde\kappa}^a\,, \nonumber \\
     & \delta_{\tilde\kappa} \omega^{ab} = 0 \,, \nonumber \\
    & \delta_{\tilde\kappa} b = 0 \,, \nonumber \\
     & \delta_{\tilde\kappa} V^a = 0 \,.
\end{align}
We remark that, under \eqref{altertrasf}, and differently from \eqref{gaugetrconf}, the fields  do not transform into the spacetime vielbein.
Consequently, we find
\begin{align}
    & \delta_{\tilde\kappa} R^{ab} = 0 \,, \nonumber \\
    & \delta_{\tilde\kappa} \mathbb{T}^a = 0 \,, \nonumber \\
    & \delta_{\tilde\kappa} \mathbb{C}^a = R^{ab} {\tilde\kappa}_b - \mathring{\mathbb{G}} {\tilde\kappa}^a= \mathbb{W}^{ab} {\tilde\kappa}_b -  \mathbb{G} {\tilde\kappa}^a \,, \nonumber \\
    & \delta_{\tilde\kappa} \mathring{\mathbb{G}} = 0 \,, 
\end{align}
from which it follows, for the field strengths appearing in the Lagrangian,
\begin{subequations}\begin{align} 
  \quad \delta_{\tilde\kappa} \mathbb{G}=&\mathcal{D}{\tilde\kappa}_a\wedge V^a\,,\\
     \delta_{\tilde\kappa} \mathbb{W}^{ab}=&-2V^{[a}\wedge \mathcal{D}{\tilde\kappa}^{b]}\,.
\end{align}
\end{subequations}
Applying the above transformations to the Lagrangian in its final form, eq. \eqref{LCG4conf} below, we find 
\begin{align}
\label{varnonssL}
\delta_{\tilde \kappa} \mathcal{L}^{\text{CG}} =& -8 a_2\,  \mathcal{D}\left({\tilde\kappa}^a \wedge V^b\right)\left[  {}^*\mathbb{W}_{ab}+  \eta_{ab}{}^*\mathbb{G}\right] 
\,,
\end{align}
which identically vanishes at $\mathbb{T}^a=0$ by virtue of  the torsion Bianchi identity.

Thus, in the case of vanishing conformal torsion, we get strict invariance of the Lagrangian 
\begin{align}
  \delta_{\tilde \kappa} \mathcal{L}^{\text{CG}} \equiv 0 \,.
\end{align}
Therefore,  the requirement of $\mathcal{H}_{\text{C}}$-invariance of our conformal theory can be interpreted as a ``symmetry-induced argument" for imposing the vanishing of $\mathbb{T}^a$ from the outset.

We observe, on the other hand, that if invariance under this Yang--Mills–like gauge symmetry is not imposed (as typically done in the conformal gravity literature), the theory may in principle admit a non-vanishing conformal torsion, in the sense that such a constraint no longer follows from a symmetry-based argument. 
In this case, the Yang--Mills structure of the Lagrangian is lost, although the Cartan bundle structure remains present -- with the Cartan connection subsuming the associated Ehresmann connection.

\section{The Lagrangian à la Cartan and its second-order expression}\label{The Lagrangian à la Cartan and its second-order expression}

Thus, the final expression of the conformal invariant Lagrangian à la Cartan is 
\begin{align}
\label{LCG4conf}  
\mathcal{L}^{\text{CG}} =& a_1  \left(\mathbb{W}^{ab}+2V^a\wedge S^b\right) \wedge  \left( \mathbb{W}^{cd}+2V^c\wedge S^d\right) \epsilon_{abcd} \nonumber \\
& +
  2\, a_2 \mathbb{W}^{ab} \wedge  ^*\mathbb{W}_{ab} 
+ 8 \, a_2  \mathbb{T}^a\wedge ^* \mathbb{C}_a - 4 \, a_2 \mathbb{G} \wedge ^*\mathbb{G} \,.
\end{align}
The latter perfectly agrees, for the bulk contributions, with the  combination of Lorentz plus Weyl invariants usually considered in the literature on Cartan conformal gravity is (see, e.g., \cite{Wheeler:2013ora,Attard:2015jfr}).

Let us notice that the Lagrangian \eqref{LCG4conf} has a MacDowell--Mansouri structure \cite{MacDowell:1977jt}, that is, up to a normalization constant, it has the form
\begin{align}
    \mathcal{L}^{\text{CG}} \propto \mathbb{F}^A \wedge ^* \mathbb{F}^B \eta_{AB} \,,
\end{align}
where $\mathbb{F}^A = (\mathbb{W}^{ab},\mathbb{G},\mathbb{C}^a,\mathbb{T}^a)$, and $\eta_{AB}$ is the Killing metric of $\rm{SO}(2,4)$.

\vskip 5mm

Written in components, and in terms of the dynamical fields, the spacetime Lagrangian \eqref{LCG4conf} reads
\begin{align}
\mathcal{L}^{\text{CG}} =    a_2\Biggl\{  &
    - R_{ab|cd}R^{ab|cd} + 8S^{(a|b)}\left( \mathcal{R}_{ab}-  S_{a|b} - \frac 12 \eta_{ab}S^c{}_{|c} \right)+\nonumber\\
  &+  8S^{[a|b]}\left( \mathcal{R}_{[ab]}- 2\mathcal{D}_{[a} b_{b]}  \right) + 8\mathcal{D}_{[a}b_{b]}\mathcal{D}^a b^b+\nonumber\\
  &+ 8\mathcal{D}_{[a}b_{b]}b_c \mathbb{T}^c{}_{|ab} +2 b_c b_d \mathbb{T}^c{}_{|ab}\mathbb{T}^{d|ab}+\,\nonumber\\    &-8S^{a|b}\left(  b_c \mathbb{T}^c{}_{|ab}+ \mathcal{D}^c \mathbb{T}_{a|bc} + \frac 12 \mathbb{T}_{a|cd}\mathbb{T}_b{}^{|cd} + {\mathbb{T}_{a|b}}^c {\mathbb{T}^d}_{|cd} \right)\Biggr\}\,\text{det}|V|\,d^4x+\nonumber\\
    &\hskip -8mm + d\left( a_1 \omega^{ab}\mathbb{R}^{cd}\epsilon_{abcd}+8 a_2 S^a \wedge {}^*\mathbb{T}_{a}\right) 
 \,,\label{lagfin0}
\end{align}
where the last line is a boundary term, that will not be further discussed.

It is interesting to inspect the  bulk Lagrangian in this final form, eq. \eqref{lagfin0}, and its field equations, since all the relevant features of conformal gravity in its standard Weyl form emerge neatly, together with some extra terms: The first line contains the only terms that contribute to the standard formulation of conformal gravity, at $\mathbb{T}^a=0$. They will be discussed later. 
As for the second line, it describes an extra dynamical sector: it involves the spin-1 gauge field $b_a$, with a ghost-like  kinetic term, and a term linear in $S_{[a|b]}$. $S_{[a|b]}$ is therefore a Lagrange multiplier enforcing the condition  $\mathcal{R}_{[ab]}-2\mathcal{D}_{[a} b_{b]}=
\text{funct}(\mathbb{T}^\ell)$ which, at $\mathbb{T}^a=0$, is consistent with the $\mathbb{T}^a$ Bianchi identity. $S_{[a|b]}$ is also a source for $b$, due to the $b_a$ field equation: $\mathcal{D}_a\mathcal{D}^{[a} b^{b]}=\mathcal{D}_a S^{[a |b]} + \text{funct}(\mathbb{T}^\ell)$. In the third and fourth line we collected all the terms proportional to the conformal torsion. In particular, at $\mathbb{T}^a \neq 0$, $S_{(a|b)}$ looks coupled to $b$ and to the torsion components.

As we are going to show below, at $\mathbb{T}^a = 0$, the Lagrangian simplifies dramatically once properly expressed at second-order: all the contributions in $S_{[a|b]}$ and $b_a$ exactly cancel out and one is left with the standard Weyl theory, where Weyl invariance is a global symmetry. 
This property was already observed, and expressed by the condition $b_a=0$, in, e.g.,  \cite{Kaku:1977pa,Kaku:1978nz} and \cite{tHooft:2014swy,Butter:2016qkx} -- in the latter works, in particular, this can be interpreted in terms of a \emph{dressing} operation \cite{Attard:2015jfr,Francois:2024rdm,Francois:2025jro}.
In the following, we will explicitly show how this mechanism is implemented in our first-order Lagrangian.

\paragraph{Making the Weyl symmetry global (second  part)\\} 

 At $\mathbb{T}^a=0$, up to the boundary term, the Lagrangian \eqref{lagfin0} boils down to
\begin{align}
\mathcal{L}^{\text{CG}} =    a_2\Biggl\{  &
    - R_{ab|cd}R^{ab|cd} + 8S^{(a|b)}\left(\mathcal{R}_{ab}- S_{a|b} -\frac 12 \eta_{ab}S^c{}_{|c} \right)+\nonumber\\
  &+  8S^{[a|b]}\left(\mathcal{R}_{[ab]}- 2\mathcal{D}_{L|[a} b_{b]} \right) + 8\mathcal{D}_{L|[a}b_{b]}\mathcal{D}_L^a b^b\Biggr\}\,\text{det}|V|\,d^4x
 \,.\label{lagfin}
\end{align}
The condition $\mathbb{T}^a=0$ is a constraint that allows to express the components of the Lorentz-spin connection in terms of $V^a_\mu$, $b_\mu$, and their derivatives. It can be rephrased in terms of the Lorentz torsion $\mathring{\mathbb{T}}^a\equiv \mathcal{D}_L V^a$ as the condition $\mathring{\mathbb{T}}^a= -b\wedge V^a$. This is useful because it allows us to write the contributions in $b_\mu$ as contorsion, that is:
${\omega^{ab}}_{|\mu}={\omega^{ab}}_{|\mu} (V,b)=\mathring{\omega}^{ab}{}_{|\mu}(V) +{\kappa^{ab}}_{|\mu}(b)$, where $\mathring{\omega}^{ab}{}_{|\mu}$ is the  torsionless (Levi--Civita), Lorentz-spin connection and ${\kappa^{ab}}_{|\mu}:=2 \delta_\mu^{[a} b^{b]}$ the contorsion.
With this substitution in \eqref{lagfin}, the $S^{[a|b]}$ term cancels out from the Lagrangian. 

Passing now to second-order, by varying the Lagrangian with respect to $S_{(a|b)}$, and substituting back the result in \eqref{lagfin}, we obtain
\begin{align}
    \mathcal{L}^{\text{CG}}_{\text{Weyl}} = a_2 \left(- \mathring{R}_{\mu \nu|\rho \sigma}\mathring{R}^{\mu \nu|\rho \sigma} + 2\mathring{\mathcal{R}}^{\mu \nu} \mathring{\mathcal{R}}_{\mu \nu} - \frac{1}{3} \mathring{\mathcal{R}}^2 \right) \sqrt{-g} \; d^4x \,,
\end{align}
which, for $a_2$ a positive, adimensional normalization constant, is the standard Weyl Lagrangian, where $\mathring{R}_{\mu \nu|\rho \sigma}$ is the Levi-Civita Riemann tensor, $\mathring{\mathcal{R}}_{\mu \nu}=\mathring{\mathcal{R}}_{\nu \mu}:=\mathring{R}_{\mu\,\,\,|\nu \rho}^{\,\,\,\,\rho}$ the Levi-Civita Ricci tensor, and $\mathring{\mathcal{R}}:=\mathring{\mathcal{R}}^\mu{}_\mu$ the associated Ricci scalar. \\
Moreover, let us mention that, by doing this, we obtain
\begin{align}
    {S^a}_{|a} & = \frac{1}{6} \mathring{\mathcal{R}} - b_a b^a - \mathcal{D}_{L|a} b^a = \frac{1}{6} \mathcal{R} \,, \\
    S_{(a|b)} & = \frac{1}{2} \mathring{\mathcal{R}}_{ab} + b_a b_b - \frac{1}{12} \eta_{ab} \left( 6 b_c b^c + \mathring{\mathcal{R}} \right) - \mathcal{D}_{L|(a} b_{b)} = \frac{1}{2} \left( \mathcal{R}_{(ab)} - \frac{1}{6} \eta_{ab} \mathcal{R} \right) \,,
\end{align}
which coincide with the usual expressions (at vanishing conformal torsion).

\section{Conclusions}\label{Conclusions}

In this work, we have constructed and analyzed a \emph{first-order} geometric Lagrangian for four-dimensional conformal gravity embedded, à la Cartan, in the conformal group  $\rm{SO}(2,4)$. 
Our Lagrangian à la Cartan depends on its gauge fields $\omega^{ab}, b, S^a$ and on the spacetime vielbein $V^a$, all fields being thought of as independent from each other, off-shell. It also includes a Lagrange multiplier to implement the condition of vanishing conformal torsion,
which is a condition generally assumed in standard conformal gravity. 
While not arising dynamically from the field equations of the spin connection as in the Einstein--Cartan theory, the above condition is shown to be required  by   local gauge invariance under the gauge group $\mathcal{H}_\text{C}$, à la Yang--Mills.\footnote{Let us mention that this result is still compatible with the analysis done in \cite{DAuria:2021dth}, where, indeed, such Yang--Mills-like invariance was not required, and the conformal gravitational theory constructed was different with respect to the one here constructed (in particular, the only term involved in the Lagrangian was the usual quadratic Weyl tensor, cf. also \cite{Kaku:1977pa,Antoniadis:1984br}).}
If invariance under the aforementioned Yang--Mills-type gauge symmetry is not imposed, the theory allows, in general, for a non-vanishing conformal torsion. 
In such a setting, the Yang--Mills structure of the theory is lost.

By varying this Lagrangian with respect to its gauge fields, the standard constraints traditionally imposed kinematically on the spin connection, the dilatation gauge field, the special conformal gauge field, and the Weyl tensor emerge \emph{dynamically}, revealing the underlying conformal Cartan bundle structure without assuming it \emph{a priori}. Moreover,  from our treatment, the mechanism that makes Weyl symmetry global, at second-order, is explicitly shown.  

Our results clarify the geometric status of conformal gravity as a gauge theory. 
In this respect, our first-order approach could significantly simplify the analysis of conformal gravity models more complicated than the four-dimensional one.
Natural perspectives include the extension of our approach to higher spacetime dimensions, where  higher-derivative theories of conformal gravity are much more intricate, see, for example, \cite{Anastasiou:2020mik,Boulanger:2025oli}.
The same applies to extensions to the superconformal case -- see, e.g., \cite{Butter:2014xxa,Butter:2017jqu}.

In the supersymmetric context, a fundamental tool for simplifying the analysis -- and in particular for formulating the theory in \emph{superspace} -- may be provided by the study of \emph{superspace cohomology} and of the possible cocycles that can be constructed therein, building on existing analyses such as those presented and discussed in, e.g., \cite{Fioresi:2015bta}, \cite{DAuria:1982uck,Andrianopoli:2016osu,Andrianopoli:2017itj,Cremonini:2022cdm,Cremonini:2024ddc}, and \cite{Imbimbo:2025ffw}.
This investigation may become possible in the superconformal theory through suitable extensions of the ``standard" cohomology (that is, of the standard cocycles, which would otherwise be trivial in the conformal case).

\section*{Acknowledgments}

We thank Jordan François and Rodrigo Olea for illuminating discussions. \\
L.R. is supported by the 
GrIFOS research project, funded by MUR, PNRR Young Researchers funding program, MSCA SoE, 
CUP E13C24003600006, ID SOE2024$\_$0000103.

\section*{Authors' Contribution Statement}

All authors contributed equally to the conception of the study, analysis, calculations, interpretation of results, and preparation of the manuscript.

\appendix

\section{Conformal algebra and conformal Cartan geometry}\label{conf algebra}

{The rigid conformal algebra $\mathfrak{conf}(1,3) \simeq \mathfrak{so}(2,4)$ is codified by the following commutation relations:
\begin{equation}
    \begin{split}
        & [\mathbf{J}_{cd},\mathbf{J}_{ef}] = -\frac{1}{2}(-\eta_{ce}\delta^{ab}_{fd} +\eta_{cf} \delta_{ed}^{ab} + \eta_{de} \delta^{ab}_{fc} -\eta_{df}\delta^{ab}_{ec})\mathbf{J}_{ab} = f^{[ab]}_{\quad [cd] [ef]}\mathbf{J}_{[ab]} \,,\\
        & [\mathbf{J}_{bc},\mathbf{P}_d] = \frac{1}{2}(\delta^a_b \eta_{cd} -\delta^a_c \eta_{bd})\mathbf{P}_a = f^a_{\;\; [bc] d}\mathbf{P}_a \,,\\
        & [\mathbf{J}_{bc},\mathbf{K}_d] = \frac{1}{2}(\delta^a_b \eta_{cd} -\delta^a_c \eta_{bd})\mathbf{K}_a = (f^K)^a_{\;\; [bc] d}\mathbf{K}_a \,, \\
        & [\mathbf{P}_a,\mathbf{D}] = -\delta_a^b \mathbf{P}_b = (f^{PD})^b_{\;\; a 0} \mathbf{P}_b\,, \\
        & [\mathbf{K}_a, \mathbf{D}] = \delta^b_a \mathbf{K}_b = (f^{KD})^b_{\;\; a0} \mathbf{K}_b\,,\\
        & [\mathbf{K}_a,\mathbf{P}_b] = \eta_{ab}\mathbf{D} -2\delta^{cd}_{ab}\mathbf{J}_{cd} = (f^{KP})^0_{\;\; ab}\mathbf{D} +(f^{KP})^{[cd]}_{\quad ab}\mathbf{J}_{[cd]}\,,
    \end{split}
\end{equation}
where  the index ``$0$" refers to the dilatation generator $\mathbf{D}$. The duality between the 1-form fields and the generators is given by
\begin{equation}\label{dualgen}
    \omega^{ab}(\mathbf{J}_{cd})=2\delta^{ab}_{cd}\,, \quad V^a(\mathbf{P}_b)=\delta^a_b \,, \quad S^a(\mathbf{K}_b)=\delta^a_b  \,, \quad b(\mathbf{D})= 1 \,.
\end{equation}
}

\subsection{Conformal group and conformal Cartan geometry (kinematics)}\label{appconf}

The group $\rm{SO}(2|4)$ is such that 
\begin{align}
    \rm{SO}(2|4) \supset H = H_0 \ltimes H_1 = (\rm{SO}(1,3)\times \rm{SO}(1,1)) \ltimes \mathbb{R}^{(1,3)*} = \begin{pmatrix}
    z & 0 & 0 \\
    0 & s & 0 \\
    0 & 0 & z^{-1}
    \end{pmatrix} 
    \begin{pmatrix}
    1 & r & \frac{1}{2} r r^t \\
    0 & \mathbf{1} & r^t \\
    0 & 0 & 1
    \end{pmatrix} \,,
\end{align}
where in the first matrix we have Lorentz and Weyl transformations, while in the second matrix of the product we have the special conformal transformations (see Proposition 1.6.3 on page 118 of \cite{capslovak:2009}). At the algebraic level:
\begin{align}
    \mathfrak{so}(2,4) \supset \mathfrak{b} = \mathfrak{b}_0 \oplus \mathfrak{b}_1 = \begin{pmatrix}
    \varepsilon & 0 & 0 \\
    0 & \mathsf{s} & 0 \\
    0 & 0 & - \varepsilon
    \end{pmatrix} \oplus
    \begin{pmatrix}
    0 & k & 0 \\
    0 & \mathbf{0} & k^t \\
    0 & 0 & 0
    \end{pmatrix} \,.
\end{align}
The conformal Cartan geometry (kinematics) is given by a principal bundle $P$ which is an $H$-bundle with gauge group $\mathcal{H} (P,\varpi)$ -- or gauge algebra Lie$\mathcal{H}$. Note that $\mathcal{H}$ \textit{is not} $\rm{SO}(2,4)$. The Cartan connection $\varpi \in \Omega^1(P,\mathfrak{so}(2,4))$ is a 1-form taking value into the Lie algebra $\mathfrak{so}(2,4)$ and reads
\begin{align}
    \varpi = \begin{pmatrix}
    b & S & 0 \\
    V & \omega & S^t \\
    0 & V^t & -b
    \end{pmatrix} \,.
\end{align}
The gauge transformation of $\varpi$ is
\begin{align}
    \varpi^\gamma = \gamma^{-1} \varpi \gamma + \gamma^{-1} d \gamma \,, \quad \gamma = \gamma_0 \gamma_1 \in \mathcal{H} = \mathcal{H}_0 \ltimes \mathcal{H}_1 \,.
\end{align}
In particular, explicitly we have
\begin{align}
    \gamma_0 = \begin{pmatrix}
    z & 0 & 0 \\
    0 & s & 0 \\
    0 & 0 & z^{-1}
    \end{pmatrix} \,,
\end{align}
which implies, by straightforward matrix computation,
\begin{align}
    \varpi^{\gamma_0} = \gamma_0^{-1} \varpi \gamma_0 + \gamma_0^{-1} d \gamma_0 = \begin{pmatrix}
    b & z^{-1} S s & 0 \\
    s^{-1} V z & s^{-1} \omega s & s^{-1} S^t z^{-1} \\
    0 & z V s & -b
    \end{pmatrix} + \begin{pmatrix}
    z^{-1} dz & 0 & 0 \\
    0 & s^{-1} d s & 0 \\
    0 & 0 & z dz^{-1}
    \end{pmatrix} \,,
\end{align}
and
\begin{align}
    \gamma_1 = \begin{pmatrix}
    1 & r & \frac{1}{2} r r^t \\
    0 & \mathbf{1} & r^t \\
    0 & 0 & 1
    \end{pmatrix} \,,
\end{align}
which yields
\begin{align}
    \varpi^{\gamma_1} = \gamma_1^{-1} \varpi \gamma_1 + \gamma_1^{-1} d \gamma_1 = \begin{pmatrix}
    b - r V & S - r \omega - \frac{1}{2} r r^t V^t + b r - r V r & 0 \\
    V & \omega + Vr - r^t V^t & * \\
    0 & V^t & -b + V^t r^t
    \end{pmatrix} + \begin{pmatrix}
    0 & dr & 0 \\
    0 & 0 & dr^t \\
    0 & 0 & 0
    \end{pmatrix} \,.
\end{align}
Thus, since infinitesimally we have
\begin{align}
    \delta_\lambda \varpi = D^{\varpi} \lambda = d \lambda + [\varpi, \lambda]\,, \quad \lambda \in \text{Lie}\mathcal{H} \,,
\end{align}
we can also write
\begin{align}
    \delta_{\lambda_0} \varpi = D^\varpi \lambda_0 \,,
\end{align}
and
\begin{align}
    \delta_{\lambda_1} \varpi = D^\varpi \lambda_1 = \begin{pmatrix}
    0 & d k & 0 \\
    0 & 0 & d k^t \\
    0 & 0 & 0
    \end{pmatrix} + \begin{pmatrix}
    0 & b k & S k^t \\
    0 & V k & \omega k^t \\
    0 & 0 & V^t k^t
    \end{pmatrix} - \begin{pmatrix}
    k V & k \omega & k S^t \\
    0 & k^t V^t & -b k^t \\
    0 & 0 & 0
    \end{pmatrix} \,,
\end{align}
that is, using the scalar product between $S$ and $k^t$ etc., 
\begin{align}
    \delta_{\lambda_1} \varpi = D^\varpi \lambda_1 = \begin{pmatrix}
    0 & d k & 0 \\
    0 & 0 & d k^t \\
    0 & 0 & 0
    \end{pmatrix} + \begin{pmatrix}
    - k V & b k - k \omega & 0 \\
    0 & V k - k^t V^t & \omega k^t - b k^t \\
    0 & 0 & V^t k^t
    \end{pmatrix} \,.
\end{align}
The latter exactly encodes \eqref{gaugetrconf}, which is indeed the infinitesimal special conformal part in $\mathcal{H}_1$ of the gauge transformations in $\mathcal{H}$ of the conformal group underlying conformal Cartan geometry.

\section{Constraints assumed in standard conformal gravity}\label{standard conf}

Let us enumerate in the following the constraints usually considered in ``standard" conformal gravity, there commonly assumed as kinematic conditions. 
They emerge  from requiring the Weyl tensor $\mathbb{W}^{ab}{}_{cd}$ to be in the irreducible representation with Dynkin label $(0,2)$ of the $\rm{SO}(1,3)$ group.
Since, from the  definition \eqref{W}
\begin{align}
\label{wwdef}
    {\mathbb{W}^{ab}}_{cd} = {R^{ab}}_{cd} - 2 {\delta^a}_{[c} {S^b}_{d]} + 2 {\delta^b}_{[c} {S^a}_{d]} \,,
\end{align}
the above requirement would imply:
\begin{enumerate}
 \item $\mathbb{W}^a{}_{[bcd]}=0
$, which means in intrinsic terms $\mathbb{W}^a_{\;\;b} \wedge V^b = 0$, and yields
\begin{align}
R^a{}_{[bcd]}=2\delta^a_{[c}S_{b|d]}   \,. 
\end{align}
This, together with $\mathcal{D}\mathbb{T}^{a}=R^a{}_b V^b+ db V^a$, implies
\begin{align}
\mathcal{D}_{[b}\mathbb{T}^a{}_{|cd]} + {\mathbb{T}^a}_{f|[b} {\mathbb{T}^f}_{|cd]} =R^a{}_{[bcd]}+2\delta^a_{[b}\mathcal{D}_c b_{d]}=   -2\delta^a_{[b}(S_{cd]}-\mathcal{D}_c b_{d]})= \delta^a_{[b}\mathbb{G}_{cd]}\,.
\end{align}

\item $\mathbb{W}_{[ab|cd]}=0
$, which implies
\begin{align}
R_{[ab|cd]}=0  \,.  
\end{align}
A useful observation is the following: once we impose $\mathbb{T}^a=0$, the condition $\mathbb{W}_{[ab|cd]}=0$ is just a consequence of the $\mathbb{T}^a$ Bianchi identity, multiplied by $V^a$: $0=\mathbb{W}_{ab}\wedge V^b \wedge V^a + \mathbb{G} \underbrace{V^a \wedge V_a}_{=0} \; \implies \; \mathbb{W}_{[ab|cd]}=0$.

\item 
 The condition $\mathbb{W}^{b}{}_{a|cb}=0$, which implies (for general $D$ spacetime dimensions):
\begin{equation}
S_{a|b}=\frac 1{D-2} \left(\mathcal{R}_{ab}-\frac 1{2(D-1)} g_{ab}\mathcal{R}\right)\,,
\end{equation}
where $\mathcal{R}_{ab}\equiv R_{ab|cd}\eta^{bd}$ is the Ricci tensor, $\mathcal{R}\equiv \mathcal{R}_{ab}\eta^{ab} $ the Ricci scalar.

\end{enumerate}

\bibliography{biblioconf.bib}

@article{Neeman:1978zvv,
    author = "Ne'eman, Y. and Regge, T.",
    title = "{Gravity and Supergravity as Gauge Theories on a Group Manifold}",
    reportNumber = "TAUP-656-78",
    doi = "10.1016/0370-2693(78)90058-8",
    journal = "Phys. Lett. B",
    volume = "74",
    pages = "54--56",
    year = "1978"
}

@article{Riegert,
    author = "Riegert, R. J.",
    title = "The particle content of linearized conformal gravity",
    journal = "Physics Letters",
    year = "1984",
    volume = "105A",
    number = "3",
    DOI = "https://doi.org/10.1016/0375-9601(84)90648-0"
}

@article{PhysRevD.26.934,
  title = {Counting of states in higher-derivative field theories},
  author = {Lee, S. C. and van Nieuwenhuizen, P.},
  journal = {Phys. Rev. D},
  volume = {26},
  issue = {4},
  pages = {934--937},
  numpages = {0},
  year = {1982},
  month = {Aug},
  publisher = {American Physical Society},
  doi = {10.1103/PhysRevD.26.934},
  url = {https://link.aps.org/doi/10.1103/PhysRevD.26.934}
}

@article{Neeman:1978njh,
    author = "Ne'eman, Y. and Regge, T.",
    editor = "Ruffini, R. and Verbin, Y.",
    title = "{Gauge Theory of Gravity and Supergravity on a Group Manifold}",
    reportNumber = "PRINT-78-0259 (IAS,-PRINCETON), ORO-3992-328",
    doi = "10.1007/BF02724472",
    journal = "Riv. Nuovo Cim.",
    volume = "1N5",
    pages = "1",
    year = "1978"
}

@article{Francois:2025jro,
    author = "Fran{\c{c}}ois, J. and Ravera, L.",
    title = "{Reassessing the foundations of metric-affine gravity}",
    eprint = "2505.05349",
    archivePrefix = "arXiv",
    primaryClass = "gr-qc",
    doi = "10.1140/epjc/s10052-025-14656-2",
    journal = "Eur. Phys. J. C",
    volume = "85",
    pages = "902",
    year = "2025"
}

@article{Francois:2024rdm,
    author = "Fran{\c{c}}ois, J. T. and Ravera, L.",
    title = "{Geometric Relational Framework for General-Relativistic Gauge Field Theories}",
    eprint = "2407.04043",
    archivePrefix = "arXiv",
    primaryClass = "gr-qc",
    doi = "10.1002/prop.202400149",
    journal = "Fortsch. Phys.",
    volume = "73",
    number = "1-2",
    pages = "2400149",
    year = "2025"
}

@article{Francois:2024rfh,
    author = "Fran\c{c}ois, J. and Ravera, L.",
    title = "{Cartan geometry, supergravity, and group manifold approach}",
    eprint = "2402.11376",
    archivePrefix = "arXiv",
    primaryClass = "math-ph",
    doi = "10.5817/AM2024-4-243",
    journal = "Archivum Math.",
    volume = "60",
    pages = "4",
    year = "2024"
}

@book{capslovak:2009,
	author = {Cap, A. and Slovak, J.},
	publisher = {American Mathematical Society},
	series = {Mathematical Surveys and Monographs},
	title = {Parabolic Geometries I: Background and General Theory},
	volume = {1},
	year = {2009}}

@article{Ferrara:1977mv,
    author = "Ferrara, S. and Zumino, B.",
    title = "{Structure of Conformal Supergravity}",
    reportNumber = "CERN-TH-2418",
    doi = "10.1016/0550-3213(78)90548-5",
    journal = "Nucl. Phys. B",
    volume = "134",
    pages = "301--326",
    year = "1978"
}

@article{Ferrara:1978rk,
    author = "Ferrara, S. and Grisaru, Marcus T. and van Nieuwenhuizen, P.",
    title = "{Poincare and Conformal Supergravity Models With Closed Algebras}",
    reportNumber = "CERN-TH-2467",
    doi = "10.1016/0550-3213(78)90389-9",
    journal = "Nucl. Phys. B",
    volume = "138",
    pages = "430--444",
    year = "1978"
}

@article{Kaku:1978nz,
    author = "Kaku, M. and Townsend, P. K. and van Nieuwenhuizen, P.",
    editor = "Salam, A. and Sezgin, E.",
    title = "{Properties of Conformal Supergravity}",
    reportNumber = "Print-78-0120 (CITY COLL., N.Y.)",
    doi = "10.1103/PhysRevD.17.3179",
    journal = "Phys. Rev. D",
    volume = "17",
    pages = "3179",
    year = "1978"
}

@article{DAuria:2021dth,
    author = "D'Auria, R. and Ravera, L.",
    title = "{Conformal gravity with totally antisymmetric torsion}",
    eprint = "2101.10978",
    archivePrefix = "arXiv",
    primaryClass = "hep-th",
    doi = "10.1103/PhysRevD.104.084034",
    journal = "Phys. Rev. D",
    volume = "104",
    number = "8",
    pages = "084034",
    year = "2021"
}

@article{Kaku:1977pa,
    author = "Kaku, M. and Townsend, P. K. and van Nieuwenhuizen, P.",
    title = "{Gauge Theory of the Conformal and Superconformal Group}",
    doi = "10.1016/0370-2693(77)90552-4",
    journal = "Phys. Lett. B",
    volume = "69",
    pages = "304--308",
    year = "1977"
}

@article{MacDowell:1977jt,
    author = "MacDowell, S. W. and Mansouri, F.",
    title = "{Unified Geometric Theory of Gravity and Supergravity}",
    reportNumber = "COO-3075-164",
    doi = "10.1103/PhysRevLett.38.739",
    journal = "Phys. Rev. Lett.",
    volume = "38",
    pages = "739",
    year = "1977",
    note = "[Erratum: Phys.Rev.Lett. 38, 1376 (1977)]"
}

@article{Harnad:1976yu,
    author = "Harnad, J. P. and Pettitt, R. B.",
    title = "{Gauge Theories for Space-Time Symmetries}",
    reportNumber = "Print-76-0141 (MONTREAL)",
    doi = "10.1063/1.522829",
    journal = "J. Math. Phys.",
    volume = "17",
    pages = "1827--1837",
    year = "1976"
}

@proceedings{Barut:1971bb,
    editor = "Barut, A. O. and Brittin, W. E.",
    title = "{De Sitter and Conformal Groups and their Applications. Proceedings,  13th Summer Institute for Theoretical Physics}: {Boulder, CO, USA, June 29-July 03, 1970}",
    publisher = "Colorado Associated Univ. Press",
    address = "Boulder",
    series = "Lectures in Theoretical Physics",
    volume = "13",
    year = "1971"
}

@article{Attard:2015jfr,
    author = "Attard, J. and Fran{\c{c}}ois, J. and Lazzarini, S.",
    title = "{Weyl gravity and Cartan geometry}",
    eprint = "1512.06907",
    archivePrefix = "arXiv",
    primaryClass = "gr-qc",
    doi = "10.1103/PhysRevD.93.085032",
    journal = "Phys. Rev. D",
    volume = "93",
    number = "8",
    pages = "085032",
    year = "2016"
}

@article{Wheeler:2013ora,
    author = "Wheeler, J. T.",
    title = "{Weyl gravity as general relativity}",
    eprint = "1310.0526",
    archivePrefix = "arXiv",
    primaryClass = "gr-qc",
    doi = "10.1103/PhysRevD.90.025027",
    journal = "Phys. Rev. D",
    volume = "90",
    number = "2",
    pages = "025027",
    year = "2014"
}

@article{Korzynski:2003bh,
    author = "Korzynski, M. and Lewandowski, J.",
    title = "{The Normal conformal Cartan connection and the Bach tensor}",
    eprint = "gr-qc/0301096",
    archivePrefix = "arXiv",
    doi = "10.1088/0264-9381/20/16/314",
    journal = "Class. Quant. Grav.",
    volume = "20",
    pages = "3745--3764",
    year = "2003"
}

@article{Bach:1921zdq,
    author = "Bach, R.",
    title = {{Zur Weylschen Relativit{\"a}tstheorie und der Weylschen Erweiterung des Kr{\"u}mmungstensorbegriffs}},
    doi = "10.1007/BF01378338",
    journal = "Math. Z.",
    volume = "9",
    number = "1",
    pages = "110--135",
    year = "1921"
}

@article{Boulanger:2025oli,
    author = "Boulanger, N. and Rovere, D.",
    title = "{8D conformal gravity with Einstein sector, and its relation to the Q-curvature}",
    journal = "arXiv:2511.01368 [hep-th]",
    year = "2025"
}

@article{Yang:1954ek,
    author = "Yang, C.-N. and Mills, R. L.",
    editor = "Hsu, Jong-Ping and Fine, D.",
    title = "{Conservation of Isotopic Spin and Isotopic Gauge Invariance}",
    doi = "10.1103/PhysRev.96.191",
    journal = "Phys. Rev.",
    volume = "96",
    pages = "191--195",
    year = "1954"
}

@incollection{Ehresmann1950,
	author = {Ehresmann, C. and Collectif},
	booktitle = {S\'eminaire Bourbaki : ann\'ees 1948/49 - 1949/50 - 1950/51, expos\'es 1-49},
	language = {fr},
	mrnumber = {1605161},
	note = {talk:24},
	number = {1},
	pages = {153-168},
	publisher = {Soci\'et\'e math\'ematique de France},
	series = {S\'eminaire Bourbaki},
	title = {Les connexions infinit\'esimales dans un espace fibr\'e diff\'erentiable},
	url = {http://www.numdam.org/item/SB_1948-1951__1__153_0},
	year = {1952}}

@article{CartanEspGen24,
	author = {\'{E}. Cartan},
	journal = {Bull. Sci. Math.},
	pages = {825-861},
	title = {Les r\'{e}centes g\'{e}n\'{e}ralisations de la notion d'espace},
	volume = {48},
	year = {1924}}

@article{CartanProj24,
	author = {\'{E}. Cartan},
	journal = {Bull. Soc. Math. France},
	pages = {205-241},
	title = {Sur les vari\'{e}t\'{e}s \`{a} connexion projective},
	volume = {52},
	year = {1924}}

@article{CartanConf23,
	author = {\'{E}. Cartan},
	journal = {Ann. Polon. Math.},
	pages = {171-221},
	title = {Les espaces \`{a} connexion conforme},
	volume = {2},
	year = {1923}}

@article{Ogiue,
	author = {K. Ogiue},
	journal = {Kodai Math. Sem. Rep.},
	pages = {193-224},
	title = {Theory of Conformal Connections},
	volume = {19},
	year = {1967}}

@book{Kobayashi1972,
	author = {S. Kobayashi},
	publisher = {Springer},
	title = {Transformation Groups in Differential Geometry},
	year = {1972}}

@book{Sharpe,
	author = {R. W. Sharpe},
	publisher = {Springer},
	series = {Graduate text in Mathematics},
	title = {Differential Geometry: Cartan's Generalization of Klein's Erlangen Program},
	volume = {166},
	year = {1996}}

@article{DAuria:1982uck,
    author = "D'Auria, R. and Fré, P.",
    title = "{Geometric Supergravity in d = 11 and Its Hidden Supergroup}",
    doi = "10.1016/0550-3213(82)90281-4",
    journal = "Nucl. Phys. B",
    volume = "201",
    pages = "101--140",
    year = "1982",
    note = "[Erratum: Nucl.Phys.B 206, 496 (1982)]"
}

@article{Andrianopoli:2016osu,
    author = "Andrianopoli, L. and D'Auria, R. and Ravera, L.",
    title = "{Hidden Gauge Structure of Supersymmetric Free Differential Algebras}",
    eprint = "1606.07328",
    archivePrefix = "arXiv",
    primaryClass = "hep-th",
    doi = "10.1007/JHEP08(2016)095",
    journal = "JHEP",
    volume = "08",
    pages = "095",
    year = "2016"
}

@article{Andrianopoli:2017itj,
    author = "Andrianopoli, L. and D'Auria, R. and Ravera, L.",
    title = "{More on the Hidden Symmetries of 11D Supergravity}",
    eprint = "1705.06251",
    archivePrefix = "arXiv",
    primaryClass = "hep-th",
    doi = "10.1016/j.physletb.2017.07.016",
    journal = "Phys. Lett. B",
    volume = "772",
    pages = "578--585",
    year = "2017"
}

@article{Cremonini:2022cdm,
    author = "Cremonini, C. A. and Grassi, P. A. and Noris, R. and Ravera, L.",
    title = "{Supergravities and branes from Hilbert-Poincar{\'e} series}",
    eprint = "2211.10454",
    archivePrefix = "arXiv",
    primaryClass = "hep-th",
    doi = "10.1007/JHEP12(2023)088",
    journal = "JHEP",
    volume = "12",
    pages = "088",
    year = "2023"
}

@article{Cremonini:2024ddc,
    author = "Cremonini, C. A. and Grassi, P. A. and Noris, R. and Ravera, L. and Santi, A.",
    title = "{Fermionic Spencer Cohomologies of D=11 Supergravity}",
    journal = "arXiv:2411.16869 [hep-th]",
    year = "accepted for publication in Advances in Theoretical and Mathematical Physics, 2024"
}

@article{Penrose-McCallum72,
	author = {R. Penrose and M. A. H. MacCallum},
	doi = {http://dx.doi.org/10.1016/0370-1573(73)90008-2},
	issn = {0370-1573},
	journal = {Physics Reports},
	number = {4},
	pages = {241 - 316},
	title = {Twistor theory: An approach to the quantisation of fields and space-time},
	url = {http://www.sciencedirect.com/science/article/pii/0370157373900082},
	volume = {6},
	year = {1973}}

@article{Penrose1977,
	author = {Penrose, R.},
	doi = {10.1016/0034-4877(77)90047-7},
	journal = {Rept. Math. Phys.},
	pages = {65-76},
	slaccitation = {%%CITATION = RMHPB,12,65;%%},
	title = {{The Twistor Program}},
	volume = {12},
	year = {1977}}

@article{Friedrich77,
	author = {Friedrich, H.},
	doi = {10.1007/BF00771141},
	issn = {1572-9532},
	journal = {General Relativity and Gravitation},
	number = {5},
	pages = {303--312},
	title = {Twistor connection and normal conformal Cartan connection},
	url = {http://dx.doi.org/10.1007/BF00771141},
	volume = {8},
	year = {1977}}

@article{Duval-et-al1983,
	author = {Burdet,G. and Duval,C. and Perrin,M.},
	doi = {10.1063/1.525927},
	eprint = {https://doi.org/10.1063/1.525927},
	journal = {Journal of Mathematical Physics},
	number = {7},
	pages = {1752-1760},
	title = {Cartan structures on Galilean manifolds: The chronoprojective geometry},
	url = {https://doi.org/10.1063/1.525927},
	volume = {24},
	year = {1983}}

@article{Weyl:1918ib,
    author = "Weyl, H.",
    title = "{Gravitation and electricity}",
    journal = "Sitzungsber. Preuss. Akad. Wiss. Berlin (Math. Phys. )",
    volume = "1918",
    pages = "465",
    year = "1918"
}

@article{Stelle:1977ry,
    author = "Stelle, K. S.",
    title = "{Classical Gravity with Higher Derivatives}",
    reportNumber = "Print-77-0417 (BRANDEIS)",
    doi = "10.1007/BF00760427",
    journal = "Gen. Rel. Grav.",
    volume = "9",
    pages = "353--371",
    year = "1978"
}

@article{Mannheim:2011ds,
    author = "Mannheim, P. D.",
    title = "{Making the Case for Conformal Gravity}",
    eprint = "1101.2186",
    archivePrefix = "arXiv",
    primaryClass = "hep-th",
    doi = "10.1007/s10701-011-9608-6",
    journal = "Found. Phys.",
    volume = "42",
    pages = "388--420",
    year = "2012"
}

@article{Maldacena:2011mk,
    author = "Maldacena, J.",
    title = "{Einstein Gravity from Conformal Gravity}",
    journal = "arXiv:1105.5632 [hep-th]",
    year = "2011"
}

@article{Mannheim:2005bfa,
    author = "Mannheim, P. D.",
    title = "{Alternatives to dark matter and dark energy}",
    eprint = "astro-ph/0505266",
    archivePrefix = "arXiv",
    doi = "10.1016/j.ppnp.2005.08.001",
    journal = "Prog. Part. Nucl. Phys.",
    volume = "56",
    pages = "340--445",
    year = "2006"
}

@article{Stelle:1976gc,
    author = "Stelle, K. S.",
    title = "{Renormalization of Higher Derivative Quantum Gravity}",
    reportNumber = "PRINT-76-1059 (BRANDEIS)",
    doi = "10.1103/PhysRevD.16.953",
    journal = "Phys. Rev. D",
    volume = "16",
    pages = "953--969",
    year = "1977"
}

@article{Lord:1985fdw,
    author = "Lord, E. A. and Goswami, P.",
    title = "{Gauging the Conformal Group}",
    doi = "10.1007/BF02847724",
    journal = "Pramana",
    volume = "25",
    pages = "635--640",
    year = "1985"
}

@book{Fioresi:2015bta,
    author = "Fioresi, R. and Lled{\'o}, M. A.",
    title = "{The Minkowski and Conformal Superspaces}: {The Classical and Quantum Descriptions}",
    doi = "10.1142/8972",
    isbn = "978-981-4566-33-9, 978-981-4566-35-3",
    publisher = "World Scientific",
    year = "2015"
}

@article{Polchinski:1987dy,
    author = "Polchinski, J.",
    title = "{Scale and Conformal Invariance in Quantum Field Theory}",
    reportNumber = "UTTG-22-87",
    doi = "10.1016/0550-3213(88)90179-4",
    journal = "Nucl. Phys. B",
    volume = "303",
    pages = "226--236",
    year = "1988"
}

@article{Nakayama:2013is,
    author = "Nakayama, Y.",
    title = "{Scale invariance vs conformal invariance}",
    eprint = "1302.0884",
    archivePrefix = "arXiv",
    primaryClass = "hep-th",
    reportNumber = "CALT-68-2910",
    doi = "10.1016/j.physrep.2014.12.003",
    journal = "Phys. Rept.",
    volume = "569",
    pages = "1--93",
    year = "2015"
}

@article{Coleman:1970je,
    author = "Coleman, S. R. and Jackiw, R.",
    title = "{Why dilatation generators do not generate dilatations?}",
    doi = "10.1016/0003-4916(71)90153-9",
    journal = "Annals Phys.",
    volume = "67",
    pages = "552--598",
    year = "1971"
}

@article{Callan:1970ze,
    author = "Callan, Jr., C. G. and Coleman, S. R. and Jackiw, R.",
    title = "{A New improved energy-momentum tensor}",
    doi = "10.1016/0003-4916(70)90394-5",
    journal = "Annals Phys.",
    volume = "59",
    pages = "42--73",
    year = "1970"
}

@article{Antoniadis:1984br,
    author = "Antoniadis, I. and Tsamis, N. C.",
    title = "{Weyl invariance and the cosmological constant}",
    journal = "SLAC-PUB-3297",
    month = "3",
    year = "1984"
}

@article{Anastasiou:2022ljq,
    author = "Anastasiou, G. and Araya, I. J. and Olea, R.",
    title = "{Energy functionals from Conformal Gravity}",
    eprint = "2209.02006",
    archivePrefix = "arXiv",
    primaryClass = "hep-th",
    doi = "10.1007/JHEP10(2022)123",
    journal = "JHEP",
    volume = "10",
    pages = "123",
    year = "2022"
}

@article{Corral:2021xsu,
    author = "Corral, C. and Giribet, G. and Olea, R.",
    title = "{Self-dual gravitational instantons in conformal gravity: Conserved charges and thermodynamics}",
    eprint = "2105.10574",
    archivePrefix = "arXiv",
    primaryClass = "hep-th",
    doi = "10.1103/PhysRevD.104.064026",
    journal = "Phys. Rev. D",
    volume = "104",
    number = "6",
    pages = "064026",
    year = "2021"
}

@article{Anastasiou:2016jix,
    author = "Anastasiou, G. and Olea, R.",
    title = "{From conformal to Einstein Gravity}",
    eprint = "1608.07826",
    archivePrefix = "arXiv",
    primaryClass = "hep-th",
    doi = "10.1103/PhysRevD.94.086008",
    journal = "Phys. Rev. D",
    volume = "94",
    number = "8",
    pages = "086008",
    year = "2016"
}

@article{Ferrara:1977ij,
    author = "Ferrara, S. and Kaku, M. and Townsend, P. K. and van Nieuwenhuizen, P.",
    title = "{Gauging the Graded Conformal Group with Unitary Internal Symmetries}",
    reportNumber = "IC/77/55",
    doi = "10.1016/0550-3213(77)90023-2",
    journal = "Nucl. Phys. B",
    volume = "129",
    pages = "125--134",
    year = "1977"
}

@article{Anastasiou:2020mik,
    author = "Anastasiou, G. and Araya, I. J. and Olea, R.",
    title = "{Einstein Gravity from Conformal Gravity in 6D}",
    eprint = "2010.15146",
    archivePrefix = "arXiv",
    primaryClass = "hep-th",
    doi = "10.1007/JHEP01(2021)134",
    journal = "JHEP",
    volume = "01",
    pages = "134",
    year = "2021"
}

@article{Butter:2014xxa,
    author = "Butter, D. and Kuzenko, S. M. and Novak, J. and Tartaglino-Mazzucchelli, G.",
    title = "{Conformal supergravity in five dimensions: New approach and applications}",
    eprint = "1410.8682",
    archivePrefix = "arXiv",
    primaryClass = "hep-th",
    reportNumber = "NIKHEF-2014-046",
    doi = "10.1007/JHEP02(2015)111",
    journal = "JHEP",
    volume = "02",
    pages = "111",
    year = "2015"
}

@article{Butter:2017jqu,
    author = "Butter, D. and Novak, J. and Tartaglino-Mazzucchelli, G.",
    title = "{The component structure of conformal supergravity invariants in six dimensions}",
    eprint = "1701.08163",
    archivePrefix = "arXiv",
    primaryClass = "hep-th",
    reportNumber = "NIKHEF-2017-004",
    doi = "10.1007/JHEP05(2017)133",
    journal = "JHEP",
    volume = "05",
    pages = "133",
    year = "2017"
}

@article{Imbimbo:2025ffw,
    author = "Imbimbo, C. and Porro, L.",
    title = "{One Ring to Rule Them All: A Unified Topological Framework for 4D Superconformal Anomalies}",
    journal = "arXiv:2507.16505 [hep-th]",
    year = "2025"
}

@article{Butter:2016qkx,
    author = "Butter, D. and Kuzenko, S. M. and Novak, J. and Theisen, S.",
    title = "{Invariants for minimal conformal supergravity in six dimensions}",
    eprint = "1606.02921",
    archivePrefix = "arXiv",
    primaryClass = "hep-th",
    reportNumber = "NIKHEF-2016-026",
    doi = "10.1007/JHEP12(2016)072",
    journal = "JHEP",
    volume = "12",
    pages = "072",
    year = "2016"
}

@article{tHooft:2014swy,
    author = "'t Hooft, Gerard",
    title = "{Local Conformal Symmetry: the Missing Symmetry Component for Space and Time}",
    journal = "arXiv:1410.6675 [gr-qc]",
    reportNumber = "ITP-UU-14-25, SPIN-14-19",
    year = "2014"
}

@article{Dneprov:2022jyn,
    author = "Dneprov, I. and Grigoriev, M.",
    title = "{Presymplectic BV-AKSZ formulation of conformal gravity}",
    eprint = "2208.02933",
    archivePrefix = "arXiv",
    primaryClass = "hep-th",
    doi = "10.1140/epjc/s10052-022-11082-6",
    journal = "Eur. Phys. J. C",
    volume = "83",
    number = "1",
    pages = "6",
    year = "2023"
}

@article{Dneprov:2024cvt,
    author = "Dneprov, I. and Grigoriev, M. and Gritzaenko, V.",
    title = "{Presymplectic minimal models of local gauge theories}",
    eprint = "2402.03240",
    archivePrefix = "arXiv",
    primaryClass = "hep-th",
    doi = "10.1088/1751-8121/ad65a3",
    journal = "J. Phys. A",
    volume = "57",
    number = "33",
    pages = "335402",
    year = "2024"
}

@article{Jizba:2014taa,
    author = "Jizba, P. and Kleinert, H. and Scardigli, F.",
    title = "{Inflationary cosmology from quantum Conformal Gravity}",
    eprint = "1410.8062",
    archivePrefix = "arXiv",
    primaryClass = "hep-th",
    doi = "10.1140/epjc/s10052-015-3441-6",
    journal = "Eur. Phys. J. C",
    volume = "75",
    number = "6",
    pages = "245",
    year = "2015"
}

@article{Mannheim:2010xw,
    author = "Mannheim, P. D. and O'Brien, J. G.",
    title = "{Fitting galactic rotation curves with conformal gravity and a global quadratic potential}",
    eprint = "1011.3495",
    archivePrefix = "arXiv",
    primaryClass = "astro-ph.CO",
    doi = "10.1103/PhysRevD.85.124020",
    journal = "Phys. Rev. D",
    volume = "85",
    pages = "124020",
    year = "2012"
}

@article{Liu:2012xn,
    author = "Liu, Hai-Shan and Lu, H.",
    title = "{Charged Rotating AdS Black Hole and Its Thermodynamics in Conformal Gravity}",
    eprint = "1212.6264",
    archivePrefix = "arXiv",
    primaryClass = "hep-th",
    doi = "10.1007/JHEP02(2013)139",
    journal = "JHEP",
    volume = "02",
    pages = "139",
    year = "2013"
}

@article{Lu:2012xu,
    author = "Lu, H. and Pang, Y. and Pope, C. N. and Vazquez-Poritz, Justin F.",
    title = "{AdS and Lifshitz Black Holes in Conformal and Einstein-Weyl Gravities}",
    eprint = "1204.1062",
    archivePrefix = "arXiv",
    primaryClass = "hep-th",
    reportNumber = "MIFPA-12-10",
    doi = "10.1103/PhysRevD.86.044011",
    journal = "Phys. Rev. D",
    volume = "86",
    pages = "044011",
    year = "2012"
}

@article{Gullu:2011sj,
    author = "Gullu, I. and Gurses, M. and Sisman, T. C. and Tekin, B.",
    title = "{AdS Waves as Exact Solutions to Quadratic Gravity}",
    eprint = "1102.1921",
    archivePrefix = "arXiv",
    primaryClass = "hep-th",
    doi = "10.1103/PhysRevD.83.084015",
    journal = "Phys. Rev. D",
    volume = "83",
    pages = "084015",
    year = "2011"
}

@inproceedings{Ayon-Beato:2012ukp,
    author = "Ayon-Beato, E. and Giribet, G. and Hassaine, M.",
    title = "{Critical gravity waves}",
    booktitle = "{13th Marcel Grossmann Meeting on Recent Developments in Theoretical and Experimental General Relativity, Astrophysics, and Relativistic Field Theories}",
    eprint = "1207.0475",
    archivePrefix = "arXiv",
    primaryClass = "gr-qc",
    doi = "10.1142/9789814623995_0085",
    pages = "1074--1076",
    year = "2015"
}

\end{document}